\documentclass[twocolumn,aps,prx,superscriptaddress,floatfix]{revtex4-2}
\usepackage[utf8]{inputenc}

\usepackage{amsmath,amssymb,bm,ulem,graphicx,siunitx}
\usepackage[colorlinks=true, linkcolor=blue, citecolor=blue, pdfencoding=auto]{hyperref}
\usepackage[dvipsnames]{xcolor}

\renewcommand{\Re}{{\rm Re}\,}
\newcommand{\ket}[1]{|#1\rangle}
\newcommand{\bra}[1]{\langle #1|}
\newcommand{\braket}[2]{\langle #1|#2\rangle}
\newcommand{\braOket}[3]{\langle #1|#2|#3\rangle}

\newcommand{\Qbf}{\ensuremath{\bm Q}}

\newcommand{\kbf}{\ensuremath{\bm k}}
\newcommand{\pbf}{\ensuremath{\bm p}}
\newcommand{\rbf}{\ensuremath{\bm r}}
\newcommand{\dbf}{\ensuremath{\bm d}}
\newcommand{\Gbf}{\ensuremath{\bm G}}

\newcommand{\up}{\uparrow}
\newcommand{\dn}{\downarrow}
\newcommand{\Up}{\Uparrow}

\newcommand{\qbf}{\ensuremath{\bm q}}
\newcommand{\Rbf}{\ensuremath{\bm R}}

\newcommand{\kappabf}{\ensuremath{\bm \kappa}}

\newcommand{\beginsupplement}{
        \setcounter{table}{0}
        \renewcommand{\thetable}{S\arabic{table}}
        \setcounter{figure}{0}
        \renewcommand{\thefigure}{S\arabic{figure}}
        \setcounter{equation}{0}
        \renewcommand{\theequation}{S\arabic{equation}}
        \setcounter{section}{0}
        \renewcommand{\thesection}{\Alph{section}}
        \setcounter{subsection}{0}
        \renewcommand{\thesubsection}{\arabic{subsection}}
}

\usepackage{tikz}
\usetikzlibrary{decorations.markings,arrows,arrows.meta}
\newcommand{\spinmomentumBerryflux}{
\begin{tikzpicture}[baseline,decoration={markings,mark=at position 0.5 with {\arrow{{Stealth[scale=1.35]}}}}]
\def\R{0.7}; 
\node at (-0.9*\R,1.5*\R) {\scriptsize \color{black} spin-$\uparrow$};
\node at (1.1*\R,1.5*\R) {\scriptsize \color{black} spin-$\downarrow$};
\draw[dash pattern = {on 3pt off 3pt}] (0.1*\R, 1.7*\R) -- (0.1*\R, -1.4*\R);
\node[circle, fill=white, draw, outer sep=0pt, inner sep=1.5pt] at (-\R,0.75*\R) (A) {};
\node[circle, fill=white, draw, outer sep=0pt, inner sep=1.5pt] at (-0.8*\R,-0.5*\R) (B) {};
\node[circle, fill=white, draw, outer sep=0pt, inner sep=1.5pt] at (1.2*\R,-0.5*\R) (C) {};
\node[circle, fill=white, draw, outer sep=0pt, inner sep=1.5pt] at (\R,0.75*\R) (D) {};
\draw[BrickRed,fill,opacity=0.2] (A.center) to[bend right=25] (B.center) to[bend right=0] (C.center) to[bend right=25] (D.center) to[bend right=0] (A.center); 
\draw[BrickRed,-{Stealth[scale=1.35]}] (A.center) to[bend right=25] (B.center) node[below left, xshift=0.7cm] {\scriptsize \color{black} $k+\delta k_a$}; 
\draw[BrickRed,-{Stealth[scale=1.35]}] (B.center) to[bend right=0] (C.center); 
\draw[BrickRed,-{Stealth[scale=1.35]}] (C.center) to[bend right=25] (D.center);
\draw[BrickRed,-{Stealth[scale=1.35]}] (D.center) to[bend right=0] (A.center)  node[left] {\scriptsize \color{black} $k$};
\node[rotate=-7] at (0.2*\R,0.05) {\scriptsize \color{BrickRed} $\delta k_a \cdot \mathcal{S}_{Q,k}^{\rm geom}$};
\end{tikzpicture}
}

\begin{document}
\title{Quantum-geometric dipole: a topological boost to flavor ferromagnetism in flat bands}
\author{Lei Chen}
\affiliation{Department of Physics and Astronomy, Stony Brook University, Stony Brook, New York 11794, USA}
\author{Sayed Ali Akbar Ghorashi}
\affiliation{Department of Physics and Astronomy, Stony Brook University, Stony Brook, New York 11794, USA}
\author{Jennifer Cano}
\affiliation{Department of Physics and Astronomy, Stony Brook University, Stony Brook, New York 11794, USA}
\affiliation{Center for Computational Quantum Physics, Flatiron Institute, New York, New York 10010, USA}
\author{Valentin Cr\'epel}
\affiliation{Center for Computational Quantum Physics, Flatiron Institute, New York, New York 10010, USA}
\affiliation{Department of Physics, University of Toronto, 60 St. George Street, Toronto, ON, M5S 1A7 Canada}

\begin{abstract}
Robust flavor-polarized phases are a striking hallmark of many flat-band moir\'e materials.
In this work, we trace the origin of this spontaneous polarization to a lesser-known quantum-geometric quantity: the \textit{quantum-geometric dipole}. 
Analogous to how the quantum metric governs the spatial spread of wavepackets, we show that the quantum-geometric dipole sets the characteristic size of particle-hole excitations, \textit{e.g.} magnons in a ferromagnet, which in turn boosts their gap and stiffness. Indeed, the larger the particle-hole separation, the weaker the mutual attraction, and the stronger the excitation energy. In topological bands, this energy enhancement admits a lower bound 
within the local-mode approximation, highlighting the crucial role of topology in flat-band ferromagnetism. We illustrate these effects in microscopic models, emphasizing their generality and relevance to moir\'e materials. Our results establish the quantum-geometric dipole as a predictive geometric indicator for ferromagnetism in flat bands, a crucial prerequisite for topological order. 
\end{abstract}

\maketitle

\paragraph*{Introduction --- } 

Quantum geometry provides a lens to understand contributions to macroscopic observables---such as conductivity, polarizability, or optical responses---that elude semiclassical explanations because they arise from how quantum states are ``sewn'' together~\cite{xiao2010berry,torma2023essay,yu2024quantum,liu2025quantum}. 
To quantify this sewing, a family of quantum geometric quantities has been introduced. The most useful among them share two key features: 
(\textit{i}) they are related to physical observables, \textit{e.g.} the quantum metric sets a bound on Wannier localization~\cite{marzari2012maximally} and governs the spin stiffness in flat-band superconductors~\cite{torma2022superconductivity,herzog2022many}; and ($ii$) they possess predictive power, offering material design principles when combined with topology~\cite{peotta2015superfluidity,xie2020topology,herzog2022superfluid,kwon2024quantum,onishi2024fundamental,komissarov2024quantum,verma2024instantaneous,onishi2024topological,esin2025quantum,froese2025probing,kattan2025linear,zeng2025superfluid,Xie2024}.

\begin{figure}
\centering
\includegraphics[width=\columnwidth]{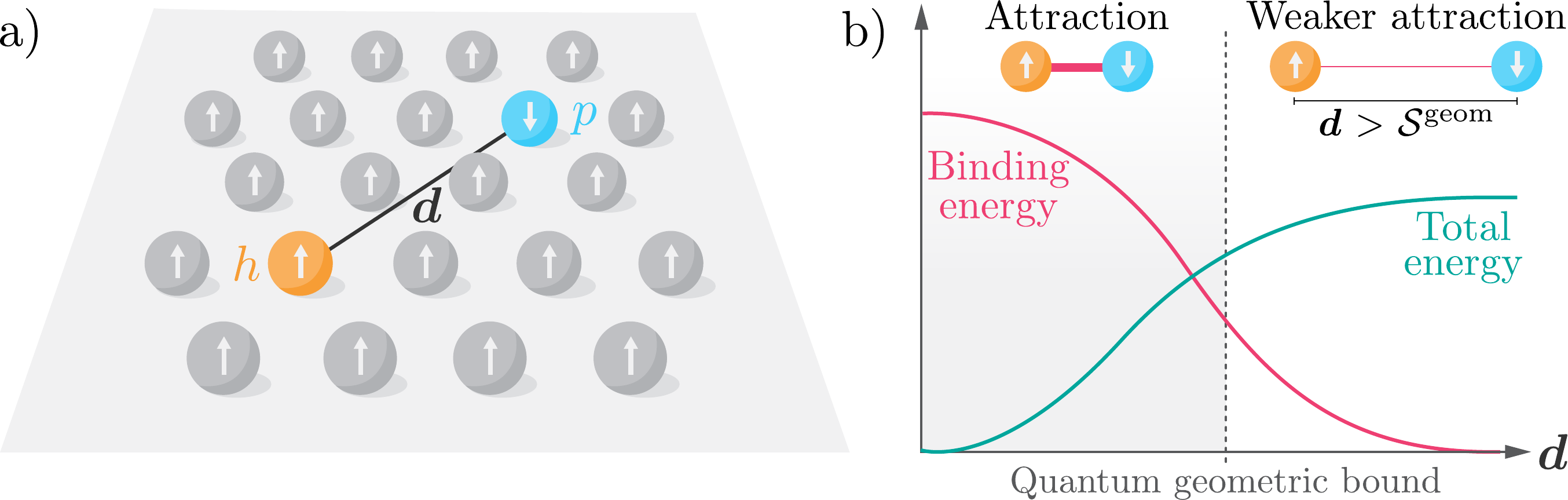}
\caption{ 
a) Schematics of the magnon dipole ${\bm d}$ measuring the distance between the $\downarrow$-particle ($p$, blue) and $\uparrow$-hole ($h$, orange) forming the flavor-flipping excitation. 
b) As the average dipole increases, the attraction between the oppositely charged $p$ and $h$ weakens, which decreases the magnon's binding energy (red) and increases its total energy (green) -- see Eq.~\ref{eq_magnoninteractiongap}. 
The quantum-geometric dipole $\mathcal{S}^{\rm geom}$ sets the typical amplitude of the magnon dipole, resulting in larger magnon energies (non-shaded area) that can be lower bounded up to $\mathcal{O}(1)$ factors by topological invariants -- see Eqs.~\ref{eq_minimumstiffness} and~\ref{eq_SMAgapbound}. 
}
\label{fig_sketch}
\end{figure}

In this work, we promote the \textit{quantum-geometric dipole} to the status of a core quantum-geometric quantity alongside the Berry curvature and quantum metric. 
Originally introduced in the context of excitonic photoresponses~\cite{von1981theory,sturman1992photovoltaic,hughes1996calculation,sipe2000second,pesin2018two,cao2021quantum}, we refine and extend the concept to show that it satisfies both criteria outlined above. 
Specifically, we first recall that it measures the typical separation between the particle ($p$) and hole ($h$) in a $ph$ excitation (Fig.~\ref{fig_sketch}a)~\cite{cao2021quantum,paiva2024shift,fertig2025many}; we then relate it to the gap and stiffness of such excitations (Fig.~\ref{fig_sketch}b); and finally show that topological invariants impose approximate lower bounds on these quantities, which become exact for flat ideal bands~\cite{estienne2023ideal,roy2014band,parameswaran2013fractional,crepel2023chiral,wang2021exact,crepel2024chiral,claassen2015position,crepel2025topologically,shi2025effects}, within the local-mode approximation.

%%%%%%%%%%%%%
%%%%%%%%%%%%%%%%%%

The relation between the $ph$ excitation energy and dipole strength admits a simple physical interpretation: as the dipole grows, the oppositely charged $p$ and $h$ become more spatially separated, their mutual attraction weakens, and the total energy of the excitation, which given by the Hartree Fock contribution and the (negative) mutual attraction energy, rises. 
The topological lower bound we derive for the $ph$ energy mirrors that of the Wannier function spread from the quantum metric~\cite{marzari1997maximally}:
just as the quantum metric measures the average spread of an electron around its center of mass, and is bounded below (up to an $\mathcal{O}(1)$ factor) by the electronic Chern number~\cite{claassen2015position,li2024constraints}; the quantum geometric dipole estimates the $ph$ distance and is similarly bounded by the difference of $p$ and $h$ Chern numbers. This, in turn, results in an approximate topological lower bound on the $ph$ energy via its dependence on the dipole. 
Our result provides a natural explanation for the widely observed tendency of topological flat bands to spontaneously polarize under interactions~\cite{zhou2021half,zhou2022isospin,de2022cascade,anderson2023programming,cai2023signatures,park2023observation,xu2023observation,zeng2023thermodynamic,han2023orbital,lu2024fractional,han2024correlated,foutty2024mapping,gao2025probing,xie2025tunable}. %\new{This close relation between ferromagnetism and topology has already been suggested by analytical calculations for specific model with exact flat bands, and gains universality through its relation with the quantum geometric dipole that we unveil.}

Quantum geometric effects are especially pronounced in flat bands, where semiclassical dynamics vanish and interactions dominate---a regime routinely realized in moir\'e~\cite{kennes2021moire,nuckolls2024microscopic,crepel2024bridging,zeng2024keldysh,guerci2024topological} and other superlattice~\cite{shi2019gate,ghorashi2023topological,ghorashi2023multilayer,krix2023patterned,gao2023untwisting,wan2023topological,zeng2024gate,seleznev2024inducing,sun2024signature,ault2025optimizing} materials, which host a wide range of correlated and topological phases~\cite{regnault2011fractional,neupert2011fractional,sheng2011fractional,crepel2024spinon,levin2009fractional,crepel2024attractive,xie2025superconductivity,stern2016fractional}.
For instance, fractionalized topological phases have recently been observed without magnetic fields in twisted MoTe$_2$~\cite{cai2023signatures,park2023observation,xu2023observation,zeng2023thermodynamic} and penta-layer graphene~\cite{lu2024fractional,xie2025tunable}. 
Crucial to their realization at experimentally accessible temperatures is a robust and spontaneous time-reversal symmetry breaking~\cite{crepel2023anomalous,morales2023pressure,repellin2020ferromagnetism,munoz2025twist,nakatsuji2025high}, enabled by extended ferromagnetic phases. Focusing on twisted MoTe$_2$, we numerically demonstrate that this robustness originates from the quantum-geometric dipole.

By clarifying its fundamental properties and crucial role in stabilizing topologically ordered phases in moir\'e materials, our work establishes the quantum-geometric dipole as a core quantum-geometric quantity. 
While our primary focus is on magnons relevant to moir\'e ferromagnets, the scope of the quantum geometric dipole extends far beyond, and could shed light on interaction-driven phenomena for excitons~\cite{verma2024geometric}, plasmons~\cite{cao2021quantum2,cao2022plasmonic}, superconductors~\cite{herzog2022many}, or multiband systems~\cite{kang2024quantum}.

\paragraph*{Quantum geometric dipole --- } The central object of our study is the magnon electric dipole, \textit{i.e.} the distance between the particle and hole forming the spin-flip excitation. 
While a similar particle–hole dipole has appeared in the context of excitonic photoresponses~\cite{von1981theory,sturman1992photovoltaic,hughes1996calculation,sipe2000second,pesin2018two,cao2021quantum,paiva2024shift,young2012first,morimoto2016topological,fei2020shift,kaplan2022twisted,chaudhary2022shift,lee2023recent,nakamura2024strongly,Haber2023}, our definition clarifies two of its key properties: 
it separates the quantum-geometric contribution from the non-universal spatial part; and it recasts the quantum-geometric part as a Berry-like flux in mixed spin–momentum space, which offers an efficient, gauge-invariant, and numerically-stable expression \textit{\`a la} Fukui-Hatsugai~\cite{fukui2005chern,fukui2007topological} (Eq.~\ref{eq_fukuihatsugai} below).

To set the stage, consider a generic band with dispersion $\varepsilon_{\kbf,\sigma}$ and periodic Bloch states $\ket{u_{k}^\sigma}$ annihilated by $c_{\kbf,\sigma}$, with $\kbf$ in the Brillouin zone (BZ) and $\sigma \in \{ \uparrow, \downarrow \}$ a flavor index. 
While our treatment extends to any band filling, we focus on the half-filled case and study magnons created over the ferromagnetic state $\ket{\hspace{-3pt}\Uparrow} = \prod_{\kbf \in {\rm BZ}} c_{\kbf,\uparrow}^\dagger \ket{\varnothing}$ with $\ket{\varnothing}$ the vacuum. 
For $\ket{\hspace{-4pt}\Uparrow}$ to be an eigenstate, we require (at least) U(1)-flavor symmetry, \textit{i.e.} that the Hamiltonian be diagonal in $\sigma$.
A generic band-projected flavor-flipping excitation with momentum $\Qbf$ can be written as
\begin{equation} \begin{split}
\ket{\psi_{\Qbf} } & = \frac{1}{\sqrt{N_{\Qbf} }} \sum_{\kbf \in {\rm BZ}} \psi_{\Qbf,\kbf} c_{\kbf+\Qbf/2,\downarrow}^\dagger c_{\kbf-\Qbf/2, \uparrow} \ket{\!\Uparrow} ,  \\
\psi_{\Qbf,\kbf} &= z_{\Qbf,\kbf} s_{\Qbf,\kbf}, \quad s_{\Qbf,\kbf} = \langle u_{\kbf+\Qbf/2}^\dn | u_{\kbf - \Qbf/2}^\up \rangle , \label{eq_genericmagnon}
\end{split} \end{equation}
with $N_{\Qbf} = \sum_{\kbf \in \text{BZ}} |\psi_{\Qbf,\kbf}|^2$ a normalization factor, and where we have decomposed the wavefunction coefficients into a gauge-invariant component $z_{\Qbf,\kbf}$ capturing the spatial structure of the magnon, and a spin-flipping form factor $s_{\Qbf,\kbf}$ originating from band projection. 

With these notations, the magnon electric dipole ${\bm d}$, defined as the average distance between the spin-$\downarrow$ particle and spin-$\uparrow$ hole, takes the form (see Ref.~\cite{paiva2024shift} or Supplemental Material (SM) Sec.~\ref{sm:qgdA})
\begin{align}
& {\bm d} = \left\langle \mathcal{S}_{\Qbf,\kbf}^{\rm spat} + \mathcal{S}_{\Qbf,\kbf}^{\rm geom}  \right\rangle_{|\psi_{\Qbf}|^2} , \quad \mathcal{S}_{\Qbf,\kbf}^{\rm spat} = i\nabla_{\kbf} \log z_{\Qbf,\kbf} ,\notag \\
& \mathcal{S}_{\Qbf,\kbf}^{\rm geom} = \mathcal{A}_{\kbf-\Qbf/2}^\uparrow - \mathcal{A}_{\kbf+\Qbf/2}^\downarrow + i \nabla_{\kbf} \log s_{\Qbf,\kbf} , \label{eq_gaugeinvariantconnection} 
\end{align}
with $\mathcal{A}_{\kbf}^\sigma = -i \braOket{u_{\kbf,\sigma}}{\nabla_{\kbf}}{u_{\kbf,\sigma}}$ the spin-resolved Berry connection, and where $\langle O_{\kbf} \rangle_\mu = \sum_{\kbf} \mu_{\kbf} O_{\kbf} / \sum_{\kbf} \mu_{\kbf}$ denotes the $\mu$-weighted average. 
Compared to the earlier definition introduced in Ref.~\cite{cao2021quantum}, we have isolated two distinct contributions to the magnon dipole. The first, dubbed the spatial dipole and denoted as $\mathcal{S}^{\rm spat}$, only depends on the $z$-coefficients and therefore originates from the structure of the dipole in real space. The second, $\mathcal{S}^{\rm geom}$, is the \textit{quantum-geometric dipole} advertised in the introduction. It is manifestly gauge-invariant~\footnote{Gauge invariance of the quantum-geometric dipole is straightforwardly checked by substituting $\ket{u_{\kbf,\sigma}} \to e^{i\theta_{\kbf,\sigma}} \ket{u_{\kbf,\sigma}}$ in Eq.~\ref{eq_gaugeinvariantconnection}.}, and depends solely on the Bloch vectors of the underlying bands. 
From Eq.~\ref{eq_gaugeinvariantconnection}, $\mathcal{S}^{\rm geom}$ encodes the discrepancy between the Berry connections of spin-$\up$ and spin-$\dn$ bands. One intuitively expects this discrepancy to be larger in a topological phase, thereby enhancing the quantum geometric dipole. The $\log s_{\Qbf,\kbf}$ term furnishes an interband compensating contribution that ensures overall gauge invariance.

To emphasize the spatial/geometric characters of the two terms in Eq.~\ref{eq_gaugeinvariantconnection}, consider the following two limits.
When the Bloch bundle is geometrically trivial, $\ket{u_{\kbf}^\sigma} = 1$, $\mathcal{S}^{\rm geom}$ vanishes and the $ph$ dipole entirely comes from the spatial part $\mathcal{S}^{\rm spat}$. 
On the other hand, the most localized magnon with center-of-mass momentum $\Qbf$, given by $\int {\rm d}^2 \Rbf \, e^{i(\Qbf\cdot \Rbf)} c_{\Rbf,\downarrow}^\dagger c_{\Rbf,\uparrow} \ket{\!\Uparrow}$, corresponds to $z_{\Qbf,\kbf} = 1$ after projection, leading to $\mathcal{S}^{\rm spat} = 0$ and a purely quantum-geometric dipole $\dbf = \langle \mathcal{S}^{\rm geom}_{\kbf} \rangle_{|\psi|^2}$.

Beyond these two limits, the geometric character of $\mathcal{S}^{\rm geom}$ becomes clear when interpreted as a Berry-like flux. To unveil it, we discretize the BZ in steps of $\delta \kbf_a = |\delta k| {\bm b}_a$ along the reciprocal lattice vectors ${\bm b}_{a=1,2}$ and approximate derivatives by finite differences to obtain at lowest order in $|\delta k|$ 
\begin{equation} \label{eq_fukuihatsugai}
\delta \kbf_a \cdot \mathcal{S}_{\Qbf,\kbf}^{\rm geom} = i \log \left[ \frac{\Lambda_{\kbf}^\up (\delta \kbf_a)}{\Lambda_{\kbf}^\dn (\delta \kbf_a)} \frac{s_{\Qbf,\kbf+\delta \kbf_a}}{s_{\Qbf,\kbf}} \right] , \; \spinmomentumBerryflux 
\end{equation} 
where $\Lambda_{\kbf}^\sigma(\qbf) = \braket{u_{\kbf+\qbf-\sigma\frac{\Qbf}{2}}^{\sigma}}{u_{\kbf-\sigma\frac{\Qbf}{2}}^{\sigma}}$ are spin-resolved form factors. 
The arrows on the accompanying sketch represent link variables~\cite{fukui2005chern,fukui2007topological} corresponding to the terms in the logarithm (for $\Qbf=0$). The closed loop they form emphasizes the gauge invariance of our discretized expression and its interpretation as a Berry flux -- or quantum geometric curvature in mixed spin-momentum.

\paragraph*{Magnon interaction energy ---} 
In the introduction, we argued that the interaction-induced magnon gap should scale with the amplitude of the magnon dipole, ${\bm d}$: a larger dipole means more separation between the spin-$\downarrow$ particle and spin-$\uparrow$ hole, which reduces their mutual attraction and thereby increases the overall magnon energy. 
To rationalize this intuition, we now compute the magnon energy and compare it with two independent approximate treatments that elucidate its connection to the dipole $\dbf$ and to the band’s underlying topology. The first is a small-$\qbf$  expansion of the form factors, which links them to quantum geometric quantities, and leads to Eq.~\ref{eq_magnoninteractiongap}. The second is a local-mode approximation where a local spin flip is projected onto the band (corresponding to $z_{\kbf}=1$ in Eq.~\ref{eq_genericmagnon}), and yields Eq.~\ref{eq_SMAgap} when combined with the first approximation.

The interaction energy of the magnon relative to the ferromagnetic state is given by  
$\Delta (\psi_{\Qbf} ) = \braOket{\psi_{\Qbf} }{H_{\rm int}}{\psi_{\Qbf}} - \braOket{\Uparrow\hspace{-5pt}}{H_{\rm int}}{\hspace{-5pt}\Uparrow}$, 
for the band-projected interaction
$H_{\rm int} =  \sum_{\substack{\qbf, \kbf ,  \pbf \\ \sigma, \tau}} \frac{v(q)}{2N_{\rm BZ}} \langle u_{\kbf+\qbf}^{\sigma}|u_{\kbf}^{\sigma}\rangle \langle u_{\pbf}^{\tau}|u_{\pbf + \qbf}^{\tau}\rangle  c_{\kbf+\qbf,\sigma}^\dagger c_{\pbf,\tau}^\dagger c_{\pbf+\qbf, \tau} c_{\kbf,\sigma}$, with $N_{\rm BZ}$ the number of points in the discretized BZ, and $v(q=||\qbf||)$ a rotation-symmetric Coulomb potential. 
Wick's theorem yields (see SM Sec.~\ref{app_magnongap})
\begin{equation} \label{eq_InteractionInTermOfLoops} 
\Delta (\psi ) = \left\langle \sum_{\qbf} \frac{v(q) |\Lambda_{\kbf}^\uparrow(\qbf)|^2}{N_{\rm BZ}} \left[ 1 - \frac{\Lambda_{\kbf}^\dn(\qbf)^*}{\Lambda_{\kbf}^{\up}(\qbf)^*} \frac{\psi_{\kbf+\qbf}}{\psi_{\kbf}} \right] \right\rangle_{\!\!|\psi|^2} \!\!\!\!\!,
\end{equation}
where we have left the $\Qbf$ dependence implicit. The total energy of the magnon also includes a kinetic term $K (\psi_{\Qbf} ) = \left\langle \varepsilon_{\kbf+\Qbf/2,\downarrow} - \varepsilon_{\kbf-\Qbf/2,\uparrow} \right\rangle_{|\psi_{\Qbf} |^2}$. The stability of the state $\ket{\hspace{-3pt}\Uparrow}$ against magnonic excitations requires that the Stoner criterion, $\min_{z_{\Qbf} } [K(\psi_{\Qbf} )+\Delta(\psi_{\Qbf})] > 0$, be fulfilled at all $\Qbf$. 
For flat bands in which the bandwidth is negligible compared to the interaction scale, we can safely ignore $K (\psi )$.

The connection between the interaction gap $\Delta(\psi)$ and the quantum-geometric dipole is brought to light by a small-$\qbf$ expansion of the form factors --- a standard approach for identifying geometric contributions to physical observables~\cite{berry1984quantal,wu2020quantum,crepel2021universal,crepel2020microscopic,abouelkomsan2023quantum,zeng2024sublattice,fang2024quantum,crepel2025efficient}. 
The zeroth order term in the expansion vanishes since the bracketed terms in Eq.~\ref{eq_InteractionInTermOfLoops} cancel out ($\Lambda_{\kbf}^{\sigma}(0)=1$). 
The linear terms also vanish due to the rotation invariance of $v(q)$. 
Hence, the leading contribution is of order $q^2$, which, by dimensional analysis, requires the introduction of a typical length scale. 
One may deduce this length scale should be the magnon's dipole by recalling that $\psi_{\kbf} \propto s_{\kbf}$ and observing that the phase of the bracketed term in Eq.~\ref{eq_InteractionInTermOfLoops} is precisely the spin-momentum Berry flux sketched in Eq.~\ref{eq_fukuihatsugai} up to the replacement $\delta \kbf_a \to \qbf$.
Confirming this intuition, the complete expansion (detailed in SM Sec.~\ref{app_magnongap}) yields
\begin{equation} \label{eq_magnoninteractiongap}
\Delta (\psi) = a^{-2}  \left\langle U (g_{\kbf} )  \left[ ||\mathcal{S}_{\kbf}^{\rm geom} + \mathcal{S}_{\kbf}^{\rm spat} ||^2 \right] \right\rangle_{|\psi|^2} , 
\end{equation}
with $a$ the lattice constant. In this calculation, the decay of the form factors for large-$\qbf$ (not captured by the small-$\qbf$ expansion) is enforced by re-exponentiating the terms proportional to the quantum metric $(g_{\kbf}^\sigma)^{ab} = \Re \braket{\partial_{k_a} u_{\kbf}^{\sigma}}{\partial_{k_b} u_{\kbf}^{\sigma}} - \mathcal{A}_{\kbf,a}^{\sigma} \mathcal{A}_{\kbf,b}^{\sigma}$.  
Considering these effective momentum cutoffs as independent of spin, we obtain the momentum-dependent interaction scale
$U (g_{\kbf}) = (4 N_{\rm BZ})^{-1} \sum_{\qbf} (qa)^2 v(q) e^{- q_a g^{ab}_{\kbf} q_b}$ 
with $g_{\kbf} = (g_{\kbf}^{\up} + g_{\kbf}^{\dn})/2$  the spin-averaged quantum metric.  
This coefficient has an intuitive interpretation: it describes the typical repulsion between two oppositely charged Gaussian wavepackets of width $g_{\kbf}$, 
the minimal spread allowed by the band's quantum geometry.
To see this, let us momentarily specialize to the two-body Coulomb potential $v(\rbf) = e^2 / (4\pi \varepsilon |\rbf|)$ and assume an isotropic quantum metric with $\tilde g = {\rm Tr} g$; then
\begin{equation}
U(g) = \frac{\pi e^2 a^2}{\varepsilon (8 \pi \tilde g)^{3/2}}  = \frac{\pi a^2}{8} \int {\rm d}^2 \rbf \, v(\rbf) |\phi(\rbf)|^4 ,
\end{equation}
with $\phi(\rbf) = \exp(- \rbf^2/2 \tilde g)/\sqrt{2\pi\tilde g}$ a normalized Gaussian wavepacket, and where $a^2$ is required by dimensionality.
Eq.~\ref{eq_magnoninteractiongap} confirms the intuitive idea that the magnon energy grows with its dipole.

Furthermore, within the {\color{blue}local}-mode approximation ({\color{blue}L}MA), the magnon gap is entirely determined by the quantum-geometric dipole. Specifically, the {\color{blue}L}MA assumes that to decrease the total dipole, the lowest energy magnons are well described by projecting local-in-space spin-flips of the form $c_{\Rbf,\downarrow}^\dagger c_{\Rbf,\uparrow} \ket{\!\Uparrow}$, \textit{i.e.} setting $z_{\kbf}=1$, so that $\psi_{\kbf} = s_{\kbf}$. This choice eliminates the spatial dipole (see discussion above Eq.~\ref{eq_fukuihatsugai}) and isolates the geometric contribution
\begin{equation}\label{eq_SMAgap}
\Delta^{\text{geom}}_{\Qbf} =  a^{-2}  \left\langle U (g_{\kbf} ) ||\mathcal{S}_{\Qbf,\kbf}^{\rm geom}  ||^2  \right\rangle_{|s|^2} ,
\end{equation}
after the small-$\qbf$ expansion.

\paragraph*{Topological boost to flat-band ferromagnetism --- } Having derived the precise relation between the magnon energy and quantum-geometric dipole in flat bands (Eq.~\ref{eq_magnoninteractiongap}), we now describe how topology underpins robust ferromagnetism in topological flat bands.

Let us start with the SU(2)-spin invariant limit $\ket{u_{\kbf,\uparrow}}=\ket{u_{\kbf,\downarrow}}$, where a gapless magnon branch with quadratic dispersion must exist~\cite{nambu1960quasi,goldstone1961field,nielsen1976count,watanabe2012unified}. 
We recover this gaplessness by noting that $\mathcal{S}_{\Qbf=0, \kbf}^{\rm geom} = 0$ for SU(2) invariant systems (see Eq.~\ref{eq_gaugeinvariantconnection}). 
Turning to Eq.~\ref{eq_magnoninteractiongap},
Taylor expanding $\Delta_{\Qbf}^{\rm geom} \simeq \rho_s |\Qbf|^2/2$ gives a magnon stiffness $\rho_s = \left\langle U (g_{\kbf}) \Omega_{\kbf}^2 \right\rangle_{\rm BZ}$ with $\Omega_{\kbf} = \nabla_{\kbf} \times \mathcal{A}_{\kbf}^{\uparrow/\downarrow}$ the Berry curvature of the band (see SM Sec.~\ref{app_magnonspinstiffness}).
When fluctuations of the quantum metric about its mean, $\bar g$, can be ignored, this formula reduces to 
\begin{equation} \label{eq_minimumstiffness}
\rho_s \simeq U(\bar g) \langle \Omega_{\kbf}^2 \rangle  \geq   U(\bar g) C^2 ,
\end{equation}
with $C = \langle \Omega_{\kbf} \rangle_{\rm BZ}$ the Chern number, and where the inequality follows from Cauchy-Schwarz. This topologically induced lower bound on the spin stiffness qualitatively agrees with previous results~\cite{wu2020quantum,chou2024topological}, and becomes exact for bands with uniform quantum metric~\cite{roy2014band}. 
It is derived assuming SU(2) symmetry, but is only non-trivial when the Chern number is non-vanishing.
Thus, it is not useful when opposite spins are related by time reversal symmetry, but yields a nontrivial bound in, \textit{e.g.} valleys of moir\'e materials away from time-reversal invariant momenta, such as the $K$ valley of twisted bilayer graphene for which analytical results have previously been derived in certain limits~\cite{kang2019strong,seo2019ferromagnetic,pons2020flat,bultinck2020mechanism,alavirad2020ferromagnetism,bultinck2020ground,bernevig2021twisted,becker2025exact,wu2020collective}.

 An important advance of our work is its applicability to spin-orbit coupled systems, where SU(2) symmetry is absent. To this end,
we now consider a flavor-symmetry lower than SU(2), for which the magnon spectrum is gapped. 
We lower bound the magnon gap by a sum of $|C_s|$ strictly positive contributions, with $C_s = C_\uparrow - C_\downarrow$ the spin Chern number of the underlying bands.
This again highlights the role of topology in flat-band ferromagnetism.  

The lower bound arises because the spin-lowering overlap $s_{\Qbf, \kbf}= \langle u_{\kbf+\Qbf/2}^{\dn}|u_{\kbf-\Qbf/2}^{\up}\rangle$ 
is a complex function in $\kbf$ whose phase needs to increase by $2\pi C_s$ as $\kbf$ winds around the BZ. 
It hence possesses at least $|C_s|$ vortices at momenta $\{\kbf_{v}\}$ with vorticity $\tau = {\rm sign} \, C_s$ (counted without multiplicities), and can be factorized as $s_{\Qbf,\kbf} = \tilde{s}_{\Qbf,\kbf} \prod_v ( k^\tau -  k_{v}^\tau ) $ with $k^\tau = k_x + i \tau k_y$ and $\tilde{s}$ a complex function with no winding. 
Near these vortices, the quantum-geometric dipole in Eq.~\ref{eq_gaugeinvariantconnection} is dominated by the singular connection $i\nabla \log s_{\Qbf}$~\cite{paiva2024shift}, which diverges as $||S_{\Qbf}^{\rm geom}|| \sim |s_{\Qbf}|^{-1}$ when $s_{\Qbf} \to 0$. 
Isolating these singular contributions in Eq.~\ref{eq_SMAgap}, we  write $\Delta_{\Qbf}^\text{geom} = \Delta_{\Qbf}^\text{bulk} + \Delta_{\Qbf}^\text{topo}$, where $\Delta_{\Qbf}^\text{topo} = \sum_v U(g_{\kbf_v})$ contains the contributions from the vortices, while $\Delta_{\Qbf}^\text{bulk} = [\sum_{\kbf \neq \kbf_v} U(g_{\kbf}) ||s_{\kbf} \mathcal{S}_{\kbf}^{\rm geom} ||^2 ] / [\sum_{\kbf} a^2 |s_{\kbf}|^2]$ gathers the contribution from all momenta away from the vortices. The positivity of the bulk part gives the topological bound
\begin{equation} \label{eq_SMAgapbound}
\Delta_{\Qbf}^\text{geom} = \Delta_{\Qbf}^\text{bulk} + \Delta_{\Qbf}^\text{topo} \geq \Delta_{\Qbf}^\text{topo} \simeq |C_s| U(\bar{g}) , 
\end{equation}
where the second (approximate) equality holds under the assumption of a near-uniform quantum metric. 
 Note that this bound is not tight: the mean quantum geometric dipole is of the order of a few times the lattice constant $a$, leading to a bulk contribution of the order of $\Delta_{\Qbf}^\text{bulk} \sim U(\bar{g})$, which is similar, up to a non-universal multiplicative factor, to the topological term. Nonetheless, it demonstrates that a vanishing gap is forbidden in the topological phase.

 Although our derivation of the lower bound assumes a homogeneous quantum geometry $g_{\kbf}=\bar{g}$, the numerical results we will show demonstrate that even in the inhomogeneous case, where 
Eq.~\ref{eq_SMAgapbound} no longer provides a strict lower bound, Eqs.~\ref{eq_magnoninteractiongap} and \ref{eq_SMAgap}
% \sout{the result can no longer be viewed as a strict lower bound,the formulas we derive }
still provide an excellent approximation to the computed gap.

The approximate topological bounds derived for the magnon stiffness (Eq.~\ref{eq_minimumstiffness}) and gap (Eq.~\ref{eq_SMAgapbound}) show the fundamental role of topology on the stability of ferromagnetism in flat bands.
They arose from the tight connection between the quantum-geometric dipole and the magnon spectrum (Eq.~\ref{eq_magnoninteractiongap}).
They are consistent with the familiar structure of spin excitations in topologically trivial models, such as the Hubbard model, where a spin-flip can be localized to a single point, which forces the quantum-geometric dipole to vanish and leads to a vanishing gap. This gaplessness in the flat-band limit implies that any additional term in the Hamiltonian (\textit{e.g.} super-exchange) can dominate, favoring different spin-ordering at low temperature (\textit{e.g.} antiferromagnetic).

 %%%%%%%%%%%%%
\begin{figure}
\centering
\includegraphics[width=1.0\columnwidth]{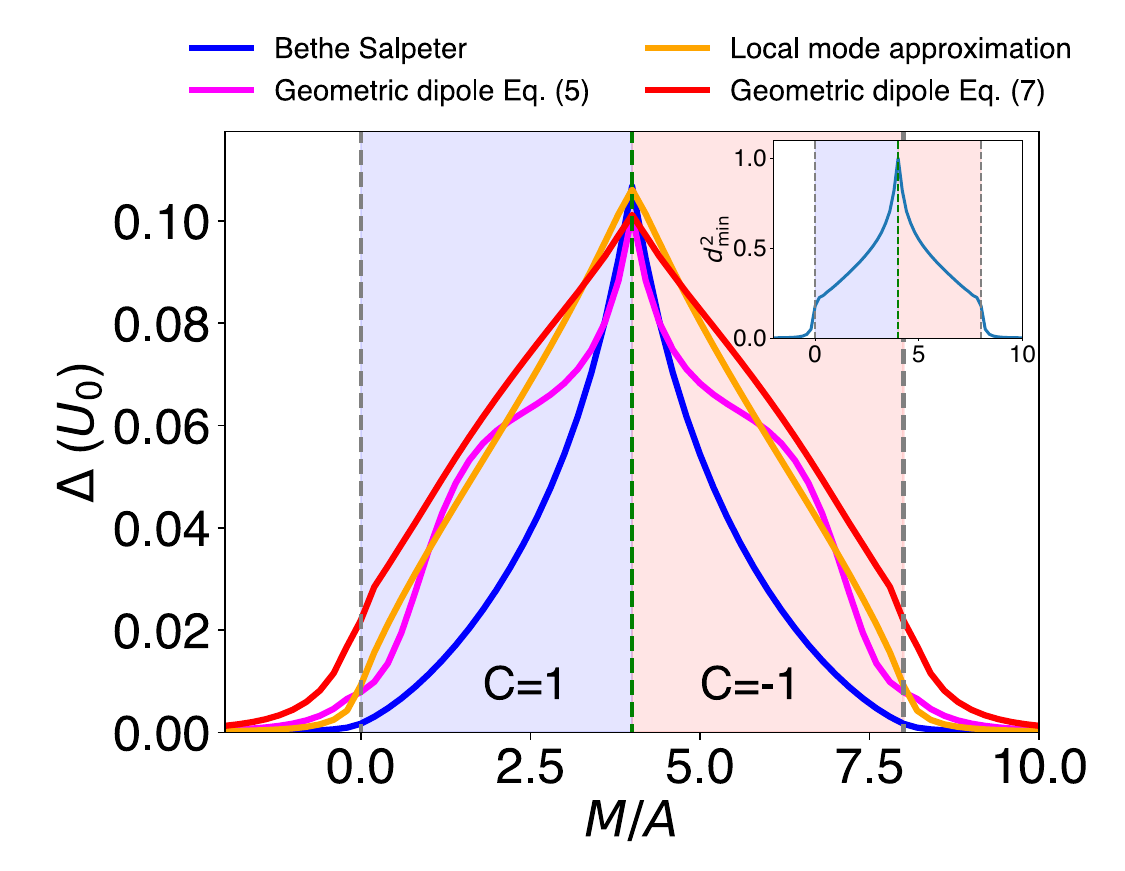}
\caption{Magnon gap versus $M/A$ for the 2D BHZ model, shown in units of $U_0$. The magnon gap is plotted using three different methods: the Bethe–Salpeter analysis (SM Eq.~\ref{eq:bs}  blue); the local-mode approximation, with $z_{\kbf} = 1$ (SM Eq.~\ref{eq:sma_s}, orange); and the geometric contribution formulas (Eq.~\ref{eq_magnoninteractiongap}, magenta; Eq.~\ref{eq_SMAgap}, red).  The shaded blue (red) regions correspond to topological phases with Chern number $C=1$ ($C=-1$) for spin-$\uparrow$.
The inset shows $d_{\rm min}^2 =  \langle ||\mathcal{S}^{\mathrm{geom}}_{\kbf}||^2 \rangle_{|s|^2}$, the average squared quantum-geometric dipole. \\ \textit{Parameters:} $A = 1$, $B = 1$, and $r_{\xi} = 0.1$.
}
\label{fig:bhz}
\end{figure}
%%%%%%%%%%%%%%%%%%

\begin{figure}
\centering
\includegraphics[width=1.0\columnwidth]{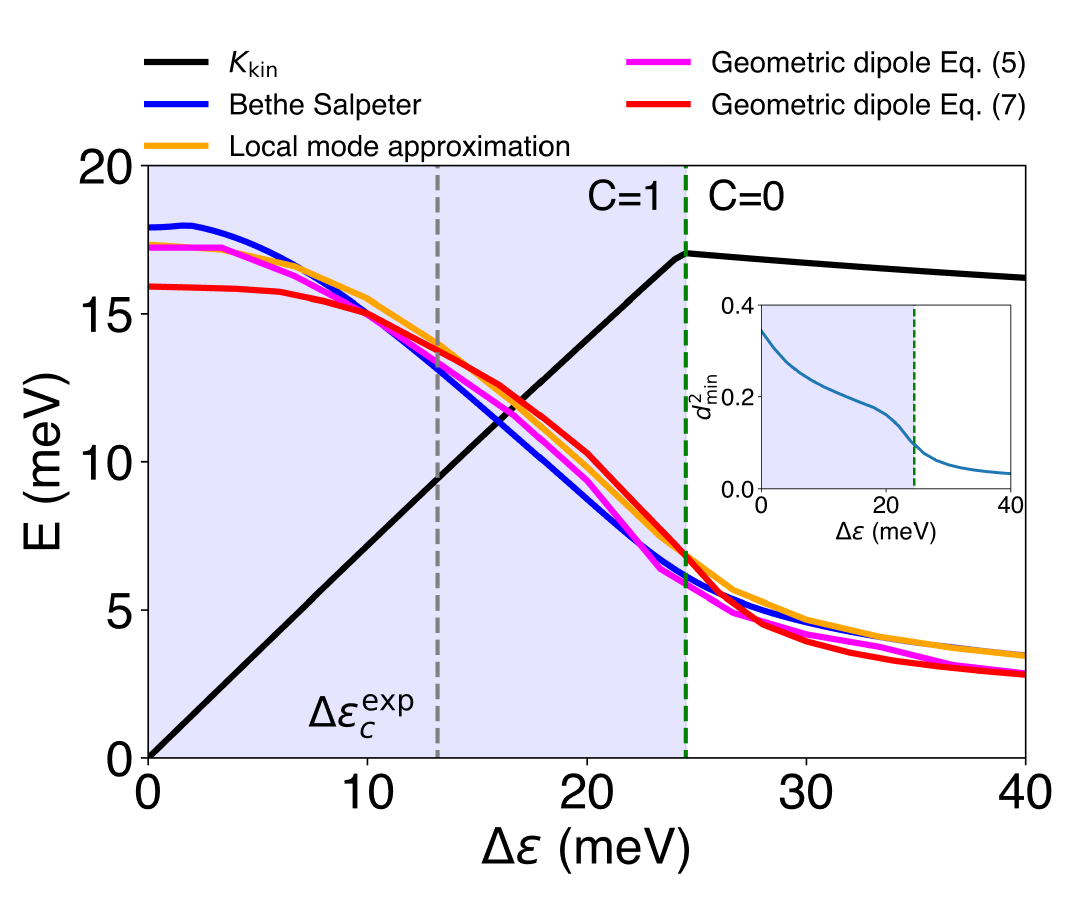}
\caption{
Same as Fig.~\ref{fig:bhz} for the continuum model of $\theta$-twisted bilayer MoTe$_2$. 
Gray dashed line marks the transition from polarized to unpolarized phases, $\Delta \varepsilon_{c}^{\rm exp} \approx 13~{\rm meV}$, extracted from the MCD measurements of Ref.~\cite{cai2023signatures}. 
Using a relaxed Stoner criterion, our theory estimates the transition point to occur at around $\Delta \varepsilon_{c}^{\rm th} \approx 16-17~{\rm meV}$, corresponding to the crossing points between the maximal kinetic energy gain $K_{\rm kin} =  \max(|\varepsilon_{\kbf, \downarrow} - \varepsilon_{\kbf, \up}|)$ (black line) and the magnon interaction energy (colored lines). 
\\ \textit{Parameters:} $\theta=3.7^{\circ}$, $\xi=30~\mathrm{nm}$, $\epsilon=7$, and $\xi_0=0.7~\mathrm{nm}$.
}
\label{fig:mote2}
\end{figure}

\paragraph*{Microscopic verification and relevance --- }
We now substantiate our theory by studying two microscopic models. We focus on systems that break SU(2) but preserve U(1) spin symmetry and compute the magnon gap at $\Qbf=0$, where the magnon spectrum attains its minimum. 
To assess the validity of previous approximations, we compute in Figs.~\ref{fig:bhz} and~\ref{fig:mote2} the magnon energy of two different models using various methods: 
% of decreasing rigor: 
the Bethe–Salpeter analysis (SM Eq.~\ref{eq:bs}, blue); the LMA before small-$\qbf$ expansion (Eq.~\ref{eq_InteractionInTermOfLoops}, orange); and the geometric contribution formulas  with spatial contribution (Eq.~\ref{eq_magnoninteractiongap}, magenta) and without spatial contribution (Eq.~\ref{eq_SMAgap}, red).
% (Eq.~\ref{eq_SMAgap}, red), with $z_{\kbf} = 1$ in the latter two.
To highlight the quantitative relationship between these energies and the quantum-geometric dipole (Fig.~\ref{fig_sketch}), we also evaluate $d_{\rm min}^2 =  \langle ||\mathcal{S}^{\mathrm{geom}}_{\kbf}||^2 \rangle_{|s|^2}$ (shown in insets). 
These quantities are presented for two representative models. First, we consider the 2D BHZ model~\cite{bernevig2006quantum} and show that the magnon gap is strong in the topological regime and sharply decreases to a near-zero value upon entering the trivial regime, which qualitatively corroborates the effect of topology identified in Eq.~\ref{eq_SMAgapbound}. 
Next, we apply our framework to twisted bilayer $\mathrm{MoTe_2}$, where spin-valley polarized phases have been observed~\cite{anderson2023programming,cai2023signatures}. At filling $\nu=1$, our coarse Stoner criterion predicts a transition to an unpolarized phase at an interlayer displacement field that closely matches experimental observations, highlighting the predictive power of our method.

\underline{$\bullet$ Lattice toy model:} 
The noninteracting part of the 2D BHZ model reads $H_0 = \mathrm{diag}[\mathcal{H}^{\uparrow}, \mathcal{H}^{\downarrow}]$, where
$\mathcal{H}^{\sigma} = \big(M - 2B(2 - \cos k_x - \cos k_y)\big) \sigma_z + A \sigma \sin k_x \sigma_x + A \sin k_y \sigma_y$. 
Choosing $A=B=1$, the model is symmetric around $M=4$ and undergoes topological phase transitions at $M = 0, 4$ and $8$ at which the Chern number shifts as $C_\uparrow = 0 \to 1 \to -1 \to 0$ (see SM Sec.~\ref{app_models}).
We then project the screened Coulomb interaction, $v(\qbf) = \frac{U_0}{r_{\xi} |\qbf|} \tanh(r_{\xi} |\qbf|)$, onto the lower band and neglect the kinetic energy to isolate interaction effects.
Fig.~\ref{fig:bhz} shows the magnon gap computed using all three methods at $r_{\xi}=0.1$. 

Two striking behaviors emerge from this plot. 
First, the average squared geometric dipole $d_{\rm min}^2$ evolves with the same trend as the magnon gap. 
This supports the physical picture developed in Fig.~\ref{fig_sketch} and formalized in Eq.~\ref{eq_SMAgap} that a greater dipole, or spatial separation between the spin-$\up$ and spin-$\dn$ forming the magnon, correlates with the ferromagnetic gap. 
Second, we notice that the ferromagnetic gap is finite across the topological region of the phase diagram, peaking near the $C=1 \to -1$ transition at $M = 4$, but sharply diminishes upon entering the trivial phase. 
This underscores the crucial role of topology in stabilizing ferromagnetic order by enhancing the average quantum-geometric dipole, as explained in Eq.~\ref{eq_SMAgapbound} and surrounding text.

%%%%%%%%%%%%%
%%%%%%%%%%%%%%%%%%

\underline{$\bullet$ Relevance to moir\'e materials:} 
To explore the interplay between topology and interaction in realistic systems, we turn to twisted bilayer $\mathrm{MoTe_2}$, where experiments have identified a broad, cone-shaped region of spin-valley polarization in the phase diagram spanned by electron filling and displacement field~\cite{anderson2023programming}.
A stable spin-valley polarized phase is a prerequisite for realizing exotic correlated states~\cite{crepel2023anomalous,qiu2025topological}, in particular for the fractional Chern insulator recently observed in this system~\cite{cai2023signatures,gonccalves2025spinless}.

We focus on the case of filling $\nu = 1$ at a twist angle of $\theta=3.7^\circ$, using the continuum moir\'e Hamiltonian (details in SM Sec.~\ref{app_models})~\cite{wu2019topological,devakul2021magic,Wang2024,Jia2024}.
An important tunable parameter is the interlayer displacement field $D$, which creates an interlayer potential difference $\Delta \varepsilon = \frac{eD}{\epsilon \epsilon_0} \xi_0$ with $\xi_0$ the interlayer spacing and $\epsilon$ the dielectric constant. 
The long-range Coulomb interaction is screened by dual metal gates, resulting in an effective potential of the form $v(\qbf) = \frac{e^2}{2 \epsilon_0 \epsilon} \frac{\tanh(\xi |\qbf|)}{ |\qbf|}$, where $\xi$ is the distance from the bilayer to the gates. 
We adopt representative experimental parameters: $\xi=30~\mathrm{nm}$, $\epsilon=7$, and $\xi_0=0.7~\mathrm{nm}$.

Fig.~\ref{fig:mote2} presents the magnon gap versus interlayer displacement potential, and demonstrates a remarkable agreement among the four methods discussed above.
Similar to the BHZ model, the difference between the topological and trivial regimes is stark: the magnon interaction energy drops to less than one-third of its initial value at $\Delta \varepsilon = 0$ upon crossing the topological transition indicated by the vertical green dashed line.

Experimentally, a transition from a ferromagnetic to an unpolarized phase is observed as a function of increasing displacement field, which we estimate around $\Delta \varepsilon_c^{\rm exp} \approx13~$meV (gray dashed line) using Ref.~\cite{cai2023signatures}'s data. 
To evaluate this transition point theoretically, we employ a relaxed Stoner criterion by approximating the kinetic contribution to the magnon energy as $K_{\rm kin} =  \max(|\varepsilon_{\kbf, \downarrow} - \varepsilon_{\kbf, \up}|)$, shown as the black solid line in Fig.~\ref{fig:mote2}, an upper bound on the kinetic energy cost $K(\psi)$ defined above. 
The ferromagnetic transition is expected to occur at the intersection between the kinetic energy cost and the magnon gap. 
This occurs around $\Delta \varepsilon_c^{\rm th} = 16-17~{\rm meV}$ for all three methods used in Fig.~\ref{fig:mote2}, which closely matches $\Delta \varepsilon_c^{\rm exp}$.

These two microscopic examples underscore the central role of the quantum-geometric dipole in the emergence of ferromagnetism. They support our approximate topological lower bound, which accounts for the widely observed tendency of topological bands to spontaneously polarize under interactions. Moreover, the strong agreement between our predictions and the depolarization transition point in twisted MoTe$_2$ highlights the quantitative accuracy and predictive power of our approach. 
 The geometric dipole of the magnon should also give rise to measurable transport phenomena, analogous to excitonic drift and shift currents~\cite{cao2021quantum,ahn2022riemannian,paiva2024shift}, although these effects would manifest at different excitation energy scales -- the energy estimates in Fig.~\ref{fig:mote2} indicate the terahertz regime. 
 
 %{\color{blue} 
 Finally, we note that for pure contact interactions the Bethe–Salpeter equation factorizes, allowing the magnon spectrum to be computed directly in orbital space (SM Sec.~\ref{sm:uniform}). For toy models with only a few orbitals, this leads to drastic numerical gains. In this contact limit, and under the assumption of strictly local spin flips and/or constraints analogous to those used in Refs.~\cite{tovmasyan2016effective,herzog2022many}, one may even expect to derive analytical expressions consistent with our results. We leave such an analysis to future work.
 %}

\paragraph*{Acknowledgments --- } 
VC is indebted to C. Paiva for an insightful presentation of her results~\cite{paiva2024shift} while the manuscript was being written. 
We also acknowledge insightful conversations with S. Divic and M. Goerbig.
LC, SAAG and JC acknowledge support from the Air Force Office of Scientific Research under Grants No. FA9550-20-1-0260 and FA9550-24-1-0222.
SAAG and JC acknowledge support from the Alfred P. Sloan
Foundation through a Sloan Research Fellowship. 
 This work was performed in part at the Aspen Center for Physics, which is supported by National Science Foundation grant PHY-2210452 and by a grant from the Simons Foundation (1161654, Troyer)
The Flatiron Institute is a division of the Simons Foundation.

\bibliography{biblioTopoZ2Gap}

\onecolumngrid 
\newpage

\beginsupplement 
\section*{Supplemental Material}
\setcounter{secnumdepth}{3}
\tableofcontents

\section{Quantum geometric dipole}
\label{sm:qgdA}

In this appendix, we derive Eq.~\ref{eq_gaugeinvariantconnection} of the main text by providing the real-space representation of the magnon wavefunction (Eq.~\ref{eq_genericmagnon}) and computing the average distance between the hole and electron that form this excitation. 

\underline{Real-space representation:} 
We consider a system whose ground state is a fully polarized ferromagnet, denoted by $\ket{\Up} = \prod_{\kbf\in{\rm BZ}}c^{\dagger}_{\kbf,\up}| \varnothing\rangle$ as defined in the main text. 
Let us denote the operator annihilating a fermion of spin $\sigma$ at position $\rbf$ as $c_{\rbf, \sigma}$, whose real-space orbital is a delta function at position $\rbf$, and write $\ket{\rbf_\downarrow, \rbf_\uparrow} = c_{\rbf_\downarrow, \downarrow}^\dagger c_{\rbf_\uparrow, \uparrow}\ket{\Up}$. 
Recalling that the $c_{\kbf, \sigma}$ operators used in the main text are the fermionic operator for \textit{a specific band} corresponding to the eigenstates $e^{i \kbf \cdot \rbf} \ket{u_{\kbf}^\sigma}$, and that the anti-commutation relation between fermionic operators is equal to the scalar product of their respective orbitals, it follows that $\{c_{\rbf, \sigma} , c_{\kbf, \sigma'}^\dagger \} = \delta_{\sigma,\sigma'} \braket{\rbf}{u_{\kbf}^\sigma} e^{i \kbf \cdot \rbf}$ where $\braket{\rbf}{u_{\kbf}^\sigma}$ denotes the real-space representation of the periodic function $\ket{u_{\kbf}^\sigma}$. 
For convenience, we repeat here the generic band-projected magnon state introduced in the main text:
\begin{equation} 
\ket{\psi_{\Qbf} } = \frac{1}{\sqrt{N_{\Qbf} }} \sum_{\kbf \in {\rm BZ}} \psi_{\Qbf,\kbf} \underbrace{c_{\kbf+\Qbf/2,\downarrow}^\dagger c_{\kbf-\Qbf/2, \uparrow} \ket{\Uparrow}}_{\ket{S_{\Qbf, \kbf}^-}} ,  \quad
\psi_{\Qbf,\kbf} = z_{\Qbf,\kbf} s_{\Qbf,\kbf}, \quad s_{\Qbf,\kbf}, = \langle u_{\kbf+\Qbf/2}^\dn | u_{\kbf - \Qbf/2}^\up \rangle , \label{eq_magnon_sm}
\end{equation}
with $N_{\Qbf} = \sum_{\kbf \in \text{BZ}} |\psi_{\Qbf,\kbf}|^2$ a normalization factor. This state can be expressed in real-space as 
\begin{align}
\braket{\rbf_\downarrow, \rbf_\uparrow}{\psi_{\Qbf}} & = \sum_{\kbf \in {\rm BZ}} \frac{\psi_{\Qbf,\kbf}}{\sqrt{N_{\Qbf}}} \braOket{\Uparrow}{ c_{\rbf_\uparrow, \uparrow}^\dagger c_{\rbf_\downarrow, \downarrow} c_{\kbf+\Qbf/2,\downarrow}^\dagger c_{\kbf-\Qbf/2, \uparrow}  }{\Up} \\
& = \sum_{\kbf \in {\rm BZ}} \frac{\psi_{\Qbf,\kbf}}{\sqrt{N_{\Qbf}}} e^{i (\kbf\cdot \rbf +\Qbf\cdot \Rbf)} \braket{\rbf_\downarrow}{u_{\kbf+\frac{\Qbf}{2}}^{\downarrow}} \braket{\rbf_\uparrow}{u_{\kbf-\frac{\Qbf}{2}}^{\uparrow}}^* , \quad \Rbf=\frac{\rbf_{\up}+\rbf_{\dn}}{2}, \quad \rbf=\rbf_{\dn}-\rbf_{\up} , 
\end{align}
where we have used the anti-commutation relation discussed above, and introduced the center of mass $\Rbf$ and relative $\rbf$ coordinates.

\underline{Average particle-hole dipole:} 
Reproducing the calculation from Ref.~\cite{paiva2024shift} for completeness, we can now compute the expectation value of the relative distance as
\begin{align}
\dbf& = \braOket{\psi}{\rbf}{\psi} = \int {\rm d}^2 \rbf_\downarrow {\rm d}^2 \rbf_\up \, \braket{\psi}{\rbf_\downarrow, \rbf_\uparrow} \rbf \braket{\rbf_\downarrow, \rbf_\uparrow}{\psi} \\ 
& = \frac{1}{N} \sum_{\kbf,\kbf'} \int {\rm d}^2 \rbf_\downarrow {\rm d}^2 \rbf_\up \, \psi_{\kbf'}^* \psi_{\kbf}  \braket{u_{\kbf'+\frac{\Qbf}{2},\downarrow}}{\rbf_\downarrow} \braket{\rbf_\uparrow}{u_{\kbf'-\frac{\Qbf}{2},\uparrow}} \braket{\rbf_\downarrow}{u_{\kbf+\frac{\Qbf}{2},\downarrow}} \braket{u_{\kbf-\frac{\Qbf}{2},\uparrow}}{\rbf_\uparrow} e^{-i (\kbf' \cdot \rbf)} \rbf e^{i (\kbf \cdot \rbf)}\\ 
& = \frac{1}{N} \sum_{\kbf,\kbf'}\int {\rm d}^2 \rbf_\downarrow {\rm d}^2 \rbf_\up \, \psi_{\kbf'}^* \psi_{\kbf} \braket{u_{\kbf'+\frac{\Qbf}{2},\downarrow}}{\rbf_\downarrow} \braket{\rbf_\uparrow}{u_{\kbf'-\frac{\Qbf}{2},\uparrow}} \braket{\rbf_\downarrow}{u_{\kbf+\frac{\Qbf}{2},\downarrow}} \braket{u_{\kbf-\frac{\Qbf}{2},\uparrow}}{\rbf_\uparrow} e^{-i (\kbf' \cdot \rbf)} (-i \nabla_{\kbf} ) e^{i (\kbf \cdot \rbf)} \label{appeq_beforeIPP} \\ 
& = \frac{1}{N} \sum_{\kbf,\kbf'} \int {\rm d}^2 \rbf_\downarrow {\rm d}^2 \rbf_\up \, \psi_{\kbf'}^* \braket{u_{\kbf'+\frac{\Qbf}{2},\downarrow}}{\rbf_\downarrow} \braket{\rbf_\uparrow}{u_{\kbf'-\frac{\Qbf}{2},\uparrow}}  e^{i (\kbf \cdot \rbf - \kbf' \cdot \rbf)} (i \nabla_{\kbf} ) \left[ \psi_{\kbf} \braket{\rbf_\downarrow}{u_{\kbf+\frac{\Qbf}{2},\downarrow}} \braket{\rbf_\uparrow}{u_{\kbf-\frac{\Qbf}{2},\uparrow}}^* \right] \label{appeq_afterIPP} \\
& = \frac{i}{N} \sum_{\kbf, \kbf'} \psi_{\kbf}^* \mathcal{O}_{\kbf - \kbf'} (u_{\kbf '+\Qbf/2}^{\dn}, u_{\kbf+\Qbf/2}^{\dn}) \mathcal{O}_{\kbf' - \kbf}  (u_{\kbf -\Qbf/2}^{\up}, u_{\kbf'-\Qbf/2}^{\up}) \nabla_{\kbf} \psi_{\kbf} \notag \\
& \qquad  + \frac{i}{N} \sum_{\kbf, \kbf'}  |\psi_{\kbf} |^2 \mathcal{O}_{\kbf - \kbf'} (u_{\kbf '+\Qbf/2}^{\dn}, \nabla_{\kbf} u_{\kbf+\Qbf/2}^{\dn}) \mathcal{O}_{\kbf' - \kbf} (u_{\kbf -\Qbf/2}^{\up}, u_{\kbf'-\Qbf/2}^{\up})  \\
& \qquad  + \frac{i}{N} \sum_{\kbf, \kbf'} |\psi_{\kbf} |^2 \mathcal{O}_{\kbf - \kbf'} (u_{\kbf '+\Qbf/2}^{\dn}, u_{\kbf+\Qbf/2}^{\dn}) \mathcal{O}_{\kbf' - \kbf} ( \nabla_{\kbf} u_{\kbf -\Qbf/2}^{\up}, u_{\kbf'-\Qbf/2}^{\up}) , \notag
\end{align}
where we have used integration by parts between lines Eq.~\ref{appeq_beforeIPP} and Eq.~\ref{appeq_afterIPP}, and have also defined the overlap $\mathcal{O}_{\qbf} (\phi_1, \phi_2) = \int {\rm d}^2 \rbf \braket{\phi_2}{\rbf} e^{i \rbf \cdot \qbf} \braket{\rbf}{\phi_1}$ between any unit-cell periodic functions $\phi_{1,2}$. Because the periodic functions carry no crystal momentum, it must be that $\mathcal{O}_{\qbf} (\phi_1, \phi_2) \propto \delta_{\qbf}$. Then the resolution of the identity yields $\mathcal{O}_{\qbf} (\phi_1, \phi_2) = \delta_{\qbf} \bra{\phi_2} ( \int {\rm d}^2 \rbf \ket{\rbf} \bra{\rbf})\ket{\phi_1} =  \delta_{\qbf} \braket{\phi_2}{\phi_1}$. 
This gives
\begin{align}
\dbf & = \frac{i}{N} \sum_{\kbf} \psi_{\kbf}^* \nabla_{\kbf} \psi_{\kbf} + |\psi_{\kbf} |^2 \left[ \braket{u_{\kbf+\frac{\Qbf}{2},\downarrow}}{\nabla_{\kbf} u_{\kbf+\frac{\Qbf}{2},\downarrow}} + \braket{\nabla_{\kbf} u_{\kbf-\frac{\Qbf}{2},\uparrow}}{ u_{\kbf-\frac{\Qbf}{2},\uparrow}} \right] \label{appeq_afterreolutionidentity}\\ 
& = \frac{i}{N} \sum_{\kbf} \psi_{\kbf}^* \nabla_{\kbf} \psi_{\kbf} + |\psi_{\kbf}|^2 \left[ \braket{u_{\kbf+\frac{\Qbf}{2},\downarrow}}{\nabla_{\kbf} u_{\kbf+\frac{\Qbf}{2},\downarrow}} - \braket{u_{\kbf-\frac{\Qbf}{2},\uparrow}}{\nabla_{\kbf}  u_{\kbf-\frac{\Qbf}{2},\uparrow}} \right] \label{appeq_afternormalized}\\ 
& =\sum_{\kbf}  \frac{|\psi_{\kbf}|^2}{N} \left[ i \nabla_{\kbf} \log z_{\kbf} + i \nabla_{\kbf} \log s_{\kbf} - \mathcal{A}_{\kbf+\frac{\Qbf}{2}}^\downarrow + \mathcal{A}_{\kbf-\frac{\Qbf}{2}}^\uparrow  \right] = \sum_{\kbf}  \frac{|\psi_{\kbf}|^2}{N} \left[ \mathcal{S}^{\rm spat}_{\kbf} + \mathcal{S}^{\rm geom}_{\kbf} \right] , 
\end{align}
where we have omitted the dependence on the center of mass momentum $\Qbf$ and used the fact that Bloch vectors are normalized to go from Eq.~\ref{appeq_afterreolutionidentity} to Eq.~\ref{appeq_afternormalized}. 
This gives Eq.~\ref{eq_gaugeinvariantconnection} in the main text.

\section{Magnon gap and quantum geometric dipole} \label{app_magnongap}

This appendix contains all details relevant to the numerical simulations presented in Figs.~\ref{fig:bhz} and~\ref{fig:mote2} of the main text, and the key steps necessary to derive our main analytical results (Eqs.~\ref{eq_InteractionInTermOfLoops}~-~\ref{eq_SMAgapbound}). 
In more details 
\begin{itemize}
    \item Sec.~\ref{sm:BS} derives the Bethe-Salpeter equation for collective excitations. 
    \item Sec.~\ref{sm:SMA} details the local-mode approximation that provides an analytically tractable approximation of the lowest-lying magnon excitation. 
    \item Sec.~\ref{sm:QGD} performs the small-$\qbf$ expansion sketched in the main text and derives the explicit expression linking the magnon gap to the quantum geometric dipole (Eq.~\ref{eq_magnoninteractiongap}). 
\end{itemize}

\subsection{Bethe-Salpeter formula and spin-flip spectrum}
\label{sm:BS}

We start from a generic interacting Hamiltonian projected onto the bands of interest
\begin{equation}
H = H_{0} + H_{\rm int}, \quad H_0 = \sum_{\kbf, \sigma} \varepsilon_{\kbf, \sigma} c_{\kbf, \sigma}^{\dagger} c_{\kbf,\sigma}, \quad 
H_{\rm int} = \frac{1}{2 } \sum_{\kbf,\pbf,\qbf,\sigma\tau} W_{\kbf\pbf}^{\sigma\tau}(\qbf) c_{\kbf+\qbf,\sigma}^{\dagger} c_{\pbf\tau}^{\dagger} c_{\pbf+\qbf,\tau}c_{\kbf\sigma} 
\end{equation}
where the interaction matrix elements $W_{\kbf\pbf}^{\sigma\tau}(\qbf)$ depend on the band form factors $\Gamma_{\kbf,\sigma}(\qbf) = \langle u_{\kbf+\qbf}^{\sigma}| u_{\kbf}^{\sigma}\rangle$ and the Fourier transform of the (rotation invariant) interaction potential $v(q = ||\qbf||)$ as 
\begin{equation} \label{appeq_interactionkernel}
W^{\sigma\tau}_{\kbf\pbf}(\qbf) =  \frac{v(\qbf)}{N_{\rm BZ}} \Gamma_{\kbf,\sigma}(\qbf) \Gamma^*_{\pbf,\tau}(\qbf) = \frac{v(\qbf)}{N_{\rm BZ}} \langle u_{\kbf+\qbf}^{\sigma}| u_{\kbf}^{\sigma}\rangle \langle u_{\pbf}^{\tau}| u_{\pbf+\qbf}^{\tau}\rangle .
\end{equation}
Projecting the Hamiltonian onto the magnon basis spanned by $\ket{S_{\Qbf, \kbf}^-}$, as introduced in Eq.~\ref{eq_magnon_sm}, we have
\begin{equation}
\begin{aligned}
\mathcal{H}_{\kbf, \kbf'}(\Qbf) &= \langle S_{\Qbf, \kbf}^- | H_0 + H_{\rm int}|  S_{\Qbf, \kbf'}^- \rangle - \langle \Uparrow | H_0 + H_{\rm int} | \Uparrow\rangle \delta_{\mathbf{k},\mathbf{k'}} \\
& = (E_{\kbf_+, \dn}-E_{\kbf_-, \up}) \delta_{\kbf, \kbf'} -  W^{\up,\dn}_{\kbf_-, \kbf_+} (\kbf'-\kbf) , \quad \kbf_\pm = \kbf \pm \Qbf / 2 ,
\end{aligned}
\label{eq:Hkkprime}
\end{equation}
where the Hartree-Fock quasiparticle energies are given by
\begin{equation}
E_{\kbf,\dn} = \varepsilon_{\kbf,\dn} + \left[ \sum_{\pbf}W_{\kbf\pbf}^{\dn\up}(0) \right] , \quad
E_{\kbf, \up} = \varepsilon_{\kbf,\up} +  \left[ \sum_{\pbf} W^{\up\up}_{\kbf\pbf}(0) -  \sum_{\pbf} W_{\pbf\pbf}^{\up\up} (\kbf -\pbf) \right] , \label{appeq_hartreedisp}
\end{equation}
under the assumption that $|\Up \rangle$ is the ground state.
Note that $\mathcal{H}$ is diagonal in the center of mass momentum $\Qbf$ of the excitation due to translation invariance.  
Noticing that the $W^{\sigma, \tau}_{\pbf, \kbf}(0) = v(0)$ is independent of spin and momentum, the Hartree terms cancel, so that:
\begin{equation}
    E_{\kbf_+,\dn} - E_{\kbf_-, \up} = \varepsilon_{\kbf_+,\dn}-\varepsilon_{\kbf_-,\up}+ \sum_{\pbf} W_{\pbf\pbf}^{\up\up} (\kbf_- -\pbf) , 
\label{appeq_hartreedisp2}
\end{equation}

The magnon spectrum is obtained by diagonalizing $\mathcal{H}_{\kbf, \kbf'}(\Qbf)$, \textit{i.e.} solving the so-called Bethe-Salpeter equation
\begin{equation}
\label{eq:bs}
%{\color{blue} 
\sum_{\kbf'}
%}
\mathcal{H}_{\kbf, \kbf'}(\Qbf) {\psi}_{\Qbf, \kbf'} = \mathcal{E}(\Qbf)  \psi_{\Qbf, \kbf} ,
\end{equation}
which yields the collective excitation energies $\mathcal{E}(\Qbf)$, and provides the coefficients $\psi_{\kbf, \Qbf}$ of the corresponding magnon wavefunctions (Eq.~\ref{eq_magnon_sm}). In the flat band limit, $\epsilon_{\kbf+,\dn} = \epsilon_{\kbf-,\up}=0$, the projected Hamiltonian reduced to
\begin{equation}\label{eq:bs_fullform}
    \mathcal{H}_{\kbf, \kbf'}(\Qbf) = \left( \sum_{\pbf} W_{\pbf\pbf}^{\up\up} (\kbf_- -\pbf)\right) \delta_{\kbf,\kbf'} -W^{\up,\dn}_{\kbf_-, \kbf_+} (\kbf'-\kbf). 
\end{equation}
where the first term corresponds to the interaction-driven Hartree–Fock splitting, while the second term represents the mutual attraction between the particle and hole. 
In Figs.~\ref{fig:bhz} and~\ref{fig:mote2}, we focus on the spectrum at $\Qbf=0$, and extract the magnon gap from the lowest eigenvalue of Eq.~\ref{eq:bs} (curves labeled ``Bethe Salpeter'' in Figs.~\ref{fig:bhz} and~\ref{fig:mote2}).

\subsection{Local mode approximation}
\label{sm:SMA}

The lowest magnon energy can also be obtained by minimizing the expectation value of $\mathcal{H}$ with respect to the coefficients of the magnon wavefunction (we omit the subscript $\Qbf$ for clarity when there is no ambiguity)
\begin{equation}
\mathcal{E}_{\rm min} (\Qbf) = \min_{\{ \psi \}}  \Delta(\psi), \quad \Delta (\psi) = \frac{1}{N} \sum_{\kbf\kbf'} \psi_{\kbf}^{*} \mathcal{H}_{\kbf, \kbf'} (\Qbf) \psi_{\kbf'} , \quad N = \sum_{\kbf} |\psi_{\kbf}|^2 .
\end{equation}
Combing Eqs.~\ref{eq:Hkkprime} -- \ref{appeq_hartreedisp2} yields
$\Delta (\psi) = K + I$, where
\begin{align}
K &= \frac{1}{N} \sum_{\kbf} |\psi_{\kbf}|^2 (\varepsilon_{\kbf_+,\dn} - \varepsilon_{\kbf_-,\up}), \label{eq:K} \\
I &= \frac{1}{N} \sum_{\kbf} \left[ |\psi_{\kbf}|^2 \sum_{\pbf} W^{\up,\up}_{\pbf, \pbf}(\kbf_- - \pbf) \right] - \frac{1}{N} \sum_{\kbf\kbf'} \left[ W^{\up,\dn}_{\kbf_-, \kbf_+} (\kbf'-\kbf)  \psi_{\kbf}^{*}\psi_{\kbf'}\right] \label{eq:I}
\end{align}
and $\kbf_\pm = \kbf \pm \Qbf/2$.
Here, $K$ captures the kinetic energy from the non-interacting band structure, while all interaction effects are gathered in $I$. 
The stability of the ferromagnetic ground state against spin-flip excitations requires a Stoner-like criterion $I + K > 0$ for any choice of $\psi$ and all momenta $\Qbf$. 
In Fig.~\ref{fig:mote2} of the main text (where we only show $\Qbf = 0$), we adopt a relaxed version of this criterion by comparing $I$ with the maximal kinetic cost $K_{\rm kin} =  \max(|\varepsilon_{\kbf, \downarrow} - \varepsilon_{\kbf, \up}|)$, which provides an upper bound of $K$.

In the flat-band limit, $K=0$, and the gap is entirely given by the interaction contribution $I$. Expanding Eq.~\ref{eq:I} with the explicit form of the interaction kernel (Eq.~\ref{appeq_interactionkernel}) and splitting $\psi_{\kbf} = z_{\kbf} s_{\kbf}$ gives 
\begin{align}
\Delta (\psi) = I & = \frac{1}{N} \sum_{\kbf, \qbf} \frac{v(\qbf)}{N_{\rm BZ}} \left[  |\psi_{\kbf} \braket{u_{\kbf_- +\qbf}^{\up}}{u_{\kbf_-}^{\up}} |^2 - \psi_{\kbf}^* \psi_{\kbf+\qbf} \braket{u_{\kbf_- +\qbf}^{\up}}{u_{\kbf_-}^{\up}} \braket{u_{\kbf_+}^{\dn}}{u_{\kbf_+ + \qbf }^{\dn}}\right]  , \\
& = \frac{1}{N} \sum_{\kbf} |\psi_{\kbf}|^2 \sum_{\qbf} \frac{v(\qbf) |\braket{u_{\kbf_- +\qbf}^{\up}}{u_{\kbf_-}^{\up}}|^2}{N_{\rm BZ}} \left[ 1 - \frac{\psi_{\kbf+\qbf}}{\psi_{\kbf}} \frac{\braket{u_{\kbf_+}^{\dn}}{u_{\kbf_+ + \qbf }^{\dn}} }{\braket{u_{\kbf_-}^{\up}}{u_{\kbf_- +\qbf}^{\up}} }\right] , \\
& = \frac{1}{N} \sum_{\kbf} |\psi_{\kbf}|^2 \sum_{\qbf} \frac{v(\qbf) |\Lambda_{\kbf}^\uparrow (\qbf)|^2}{N_{\rm BZ}} \left[ 1 - \frac{\psi_{\kbf+\qbf}}{\psi_{\kbf}} \frac{ \Lambda_{\kbf}^\downarrow (\qbf)^*}{\Lambda_{\kbf}^\uparrow (\qbf)^*}\right] , \label{appeq_reproducesInteractionInTermOfLoops}
\end{align}
where we have used the notation $\Lambda_{\kbf}^\sigma(\qbf) = \braket{u_{\kbf+\qbf-\sigma\frac{\Qbf}{2}}^{\sigma}}{u_{\kbf-\sigma\frac{\Qbf}{2}}^{\sigma}}$, with $\sigma=\pm1$ for spin-$\up/\dn$ (introduced in the main text) to match the form of Eq.~\ref{eq_InteractionInTermOfLoops}. 

The local-mode approximation discussed in the main text takes $z_{\kbf}=1$, yielding
\begin{equation}
\Delta^{\rm  LMA} (\psi)  = \frac{1}{N} \sum_{\kbf} |s_{\kbf}|^2 \sum_{\qbf} \frac{v(\qbf) |\Lambda_{\kbf}^\uparrow (\qbf)|^2}{N_{\rm BZ}} \left[ 1 - \frac{s_{\kbf+\qbf}}{s_{\kbf}} \frac{ \Lambda_{\kbf}^\downarrow (\qbf)^*}{\Lambda_{\kbf}^\uparrow (\qbf)^*}\right] ,
\label{eq:sma_s}
\end{equation}
which corresponds to the curves labeled ``local mode approximation'' in Figs.~\ref{fig:bhz} and~\ref{fig:mote2}.

\subsection{Relation between magnon gap and quantum geometric dipole}
\label{sm:QGD}

We further approximate $\Delta(\psi)$ by neglecting large-$\qbf$ processes. 
This is justified because both the form factor magnitudes and the screened Coulomb interaction decay with increasing momentum transfer $\qbf$. 
Specifically, we expand the form factors to second order in $\qbf$ to capture the small-$\qbf$ behavior, and re-exponentiate the parts that includes the quantum metric to ensure fast decay for large-$\qbf$. 
For instance, the norm of the spin-conserving form factors approximately behaves as
\begin{align}
|\braket{u_{\kbf+\qbf}^{\sigma}}{u_{\kbf}^{\sigma}}|^2 & = \braket{u_{\kbf+\qbf}^{\sigma}}{u_{\kbf}^{\sigma}} \braket{u_{\kbf}^{\sigma}}{u_{\kbf+\qbf}^{\sigma}}  \\
& \approx 1 + \frac{q_a}{2} \left( \braket{\partial_a u_{\kbf}^{\sigma}}{u_{\kbf}^{\sigma}} + \braket{u_{\kbf}^{\sigma}}{\partial_a u_{\kbf}^{\sigma}} \right) + \frac{q_a q_b}{2} (\braket{\partial_a \partial_b u_{\kbf}^{\sigma}}{u_{\kbf}^{\sigma}} + \braket{u_{\kbf}^{\sigma}}{\partial_a \partial_b u_{\kbf}^{\sigma}} - 2 \braket{\partial_a u_{\kbf}^{\sigma}}{u_{\kbf}^{\sigma}}  \braket{ u_{\kbf}^{\sigma}}{\partial_b u_{\kbf}^{\sigma}} ) \\
& = 1 - q_a q_b [ \Re \braket{\partial_a u_{\kbf}^{\sigma}}{\partial_b u_{\kbf}^{\sigma}} - \mathcal{A}_{\kbf,a}^\sigma \mathcal{A}_{\kbf,b}^\sigma  ]  =  1 - q_a q_b g_{\kbf, ab}^\sigma \label{appeq_expansionformfactor} \\
&\approx e^{ - q_a g_{\kbf, ab}^\sigma  q_b} , 
\end{align}
with summation over repeated spatial indices implied, and where the two approximate equal signs indicate the small-$\qbf$ expansion and the re-exponentiation, respectively. 
We have introduced the quantum metric $g^\sigma$ and Berry connection $ \mathcal{A}^\sigma$ defined as
\begin{equation}
g_{\kbf,ab}^{\sigma} = \Re \braket{\partial_{k_a} u_{\kbf}^{\sigma}}{\partial_{k_b} u_{\kbf}^{\sigma}} - \mathcal{A}_{\kbf,a}^{\sigma} \mathcal{A}_{\kbf,b}^{\sigma} , \quad  \mathcal{A}_{\kbf, a}^{\sigma} = -i \langle u_{\kbf}^{\sigma}| \partial_{k_a}| u_{\kbf}^{\sigma} \rangle ,
\end{equation} 
and have used the normalization of Bloch eigenstates to derive the identities
\begin{align}
& \braket{u_{\kbf}^\sigma}{u_{\kbf}^\sigma} = 1 , \quad \partial_a \braket{u_{\kbf}^\sigma}{u_{\kbf}^\sigma} = \braket{ \partial_au_{\kbf}^\sigma}{u_{\kbf}^\sigma} + \braket{u_{\kbf}^\sigma}{\partial_a u_{\kbf}^\sigma} = 0 , \\ 
& \partial_a \partial_b \braket{u_{\kbf}^\sigma}{u_{\kbf}^\sigma} = \braket{\partial_a \partial_b u_{\kbf}^{\sigma}}{u_{\kbf}^{\sigma}} + \braket{u_{\kbf}^{\sigma}}{\partial_a \partial_b u_{\kbf}^{\sigma}} + 2 \underbrace{\Re \braket{\partial_a u_{\kbf}^{\sigma}}{\partial_b u_{\kbf}^{\sigma}}}_{=g_{\kbf,ab}^\sigma + \mathcal{A}_{\kbf,a}^\sigma \mathcal{A}_{\kbf,b}^\sigma}  = 0 .
\end{align}

To apply these two steps on $\Delta(\psi)$, we will separately consider the following two contributions 
\begin{equation}
\Delta(\psi) = \Delta_1(\psi) - \Delta_2 (\psi) , \quad \Delta_1(\psi)  = \frac{1}{N} \sum_{\qbf} \frac{v(\qbf)}{N_{\rm BZ}} \sum_{\kbf} |\psi_{\kbf} \Lambda_{\kbf}^\uparrow (\qbf)|^2 , \quad \Delta_2(\psi) = \frac{1}{N} \sum_{\qbf} \frac{v(\qbf)}{N_{\rm BZ}} \sum_{\kbf} \psi_{\kbf}^* \psi_{\kbf +\qbf} \Lambda_{\kbf}^\uparrow (\qbf) \Lambda_{\kbf}^\downarrow (\qbf)^* .
\end{equation}
The approximation of $\Delta_1(\psi)$ follows straightforwardly from our calculation on the form factors above:
\begin{equation} \label{appeq_delta1}
\Delta_1(\psi)  = \frac{1}{N} \sum_{\qbf} \frac{v(\qbf)}{N_{\rm BZ}} \sum_{\kbf} |\psi_{\kbf}|^2 | \braket{u_{\kbf_- + \qbf}^{\uparrow}}{u_{\kbf_-}^{\uparrow}} |^2 \simeq \frac{1}{N} \sum_{\kbf, \qbf} \frac{v(\qbf)}{N_{\rm BZ}} |\psi_{\kbf}|^2 e^{-q_a g_{\kbf_-, ab}^\uparrow q_b} . 
\end{equation}

Finding the approximate behavior of $\Delta_2(\psi)$ requires more caution and can be split into three steps. \underline{The first step} is to shift (without loss of generality) $\kbf \to \kbf -\qbf/2$ in the $\kbf$-integration as above, and expand the integrand
\begin{align}
& \psi_{\kbf - \qbf/2}^* \psi_{\kbf + \qbf/2} \Lambda_{\kbf - \qbf/2}^\uparrow (\qbf) \Lambda_{\kbf - \qbf/2}^\downarrow (\qbf)^* = \psi_{\kbf - \qbf/2}^* \psi_{\kbf + \qbf/2} \braket{u_{\kbf_- + \qbf/2}^\uparrow}{u_{\kbf_- - \qbf/2}^\uparrow} \braket{u_{\kbf_+ - \qbf/2}^\downarrow}{u_{\kbf_+ + \qbf/2}^\downarrow} \\ 
& \approx |\psi_{\kbf}|^2 +  \frac{|\psi_{\kbf}|^2 q_a}{2} [\partial_a \log\psi_{\kbf} - \partial_a \log\psi_{\kbf}^* + 2 i ( \mathcal{A}_{\kbf_+,a}^\downarrow - \mathcal{A}_{\kbf_-,a}^\uparrow ) ] \notag \\ 
& \qquad \qquad + \frac{ |\psi_{\kbf}|^2 q_a q_b}{4} \left[ \frac{\partial_a \partial_b \psi_{\kbf}^*}{2 \psi_{\kbf}^*} + \frac{\partial_a \partial_b \psi_{\kbf}}{2 \psi_{\kbf}} - \frac{\partial_a \psi_{\kbf} \partial_b \psi_{\kbf}^*}{|\psi_{\kbf}|^2} \right] \notag \\ 
& \qquad\qquad - \frac{ |\psi_{\kbf}|^2 q_a q_b}{2}  \left[ g_{\kbf_-, ab}^\uparrow + \mathcal{A}_{\kbf_-,a}^\uparrow \mathcal{A}_{\kbf_-,b}^{\up} + g_{\kbf_+, ab}^\downarrow + \mathcal{A}_{\kbf_+,a}^\downarrow \mathcal{A}_{\kbf_+,b}^\downarrow \right] \notag \\ 
& \qquad\qquad + i \frac{ |\psi_{\kbf}|^2 q_a q_b}{2} \left[ ( \partial_a \log \psi_{\kbf} - \partial_a \log \psi_{\kbf}^* )( \mathcal{A}_{\kbf_+,b}^\downarrow -\mathcal{A}_{\kbf_-,b}^\uparrow ) \right] + |\psi_{\kbf}|^2 q_a q_b \mathcal{A}_{\kbf_+,b}^\downarrow \mathcal{A}_{\kbf_-,b}^\uparrow . \label{appeq_lastlineuglyexpansion}
\end{align}
In Eq.~\ref{appeq_lastlineuglyexpansion}, the first line gathers the zeroth and first order terms in $\qbf$.
The second line contains all quadratic terms arising solely from the expansion of $\psi_{\kbf - \qbf/2}^* \psi_{\kbf + \qbf/2}$.
The third line includes the quadratic terms from the form factor expansion (given in Eq.~\ref{appeq_expansionformfactor}).
The final line gathers all remaining quadratic terms that result from products of first-order terms not accounted for in the previous lines.
An elegant way to regroup these terms is to realize that, with $S_{\kbf}^{\rm tot} = S_{\kbf}^{\rm geom} + S_{\kbf}^{\rm spat}$, 
\begin{align}
S_{\kbf,a}^{{\rm tot}, *} S_{\kbf,b}^{\rm tot} & = (\mathcal{A}_{\kbf_-,a}^\uparrow - \mathcal{A}_{\kbf_+,a}^\downarrow - i \partial_a \log \psi^* ) (\mathcal{A}_{\kbf_-,b}^\uparrow  - \mathcal{A}_{\kbf_+,b}^\downarrow + i \partial_b \log\psi ) \\ 
& = \mathcal{A}_{\kbf_-,a}^\uparrow \mathcal{A}_{\kbf_-,b}^\uparrow + \mathcal{A}_{\kbf_+,a}^\downarrow \mathcal{A}_{\kbf_+,b}^\downarrow + \partial_a \log \psi^* \partial_b \log \psi  - 2 \mathcal{A}_{\kbf_+,a}^\downarrow \mathcal{A}_{\kbf_-,b}^\uparrow -i ( \partial_a \log \psi_{\kbf} - \partial_a \log \psi_{\kbf}^* )( \mathcal{A}_{\kbf_+,b}^\downarrow -\mathcal{A}_{\kbf_-,b}^\uparrow ) \notag
\end{align}
contains most of the terms appearing in Eq.~\ref{appeq_lastlineuglyexpansion} such that  
\begin{align}
\psi_{\kbf - \qbf/2}^* \psi_{\kbf + \qbf/2} \Lambda_{\kbf - \qbf/2}^\uparrow (\qbf) \Lambda_{\kbf - \qbf/2}^\downarrow (\qbf)^* & \approx |\psi_{\kbf}|^2 +  \frac{|\psi_{\kbf}|^2 q_a}{2} [\partial_a \log\psi_{\kbf} - \partial_a \log\psi_{\kbf}^* + 2 i ( \mathcal{A}_{\kbf_+,a}^\downarrow - \mathcal{A}_{\kbf_-,a}^\uparrow ) ]  \label{appeq_slightlynicerexpansionofDelta2integrand} \\ 
& \qquad + \frac{q_a q_b}{8} [ \partial_a \partial_b |\psi_{\kbf}|^2 ] - \frac{ |\psi_{\kbf}|^2 q_a q_b}{2}  \left[ g_{\kbf_-, ab}^\uparrow + g_{\kbf_+, ab}^\downarrow \right] - \frac{|\psi_{\kbf}|^2 q_a q_b}{2} S_{\kbf,a}^{{\rm tot}, *} S_{\kbf,b}^{\rm tot} . \notag 
\end{align}
\underline{The second step} is to simplify this expansion by factorization of either the $\qbf$ or $\kbf$ sum in the expression of $\Delta_2(\psi)$. 
For instance, we notice that Eq.~\ref{appeq_slightlynicerexpansionofDelta2integrand} consists of products of $\qbf$ monomials and $\qbf$-independent terms, such that the sums over $\qbf$ can be factorized and lead to simplifications when we assume that a rotation-invariant Coulomb potential. In that case, we indeed have 
\begin{equation}
F_{a} \sum_{\qbf} \frac{v(q)}{N_{\rm BZ}} q_a = 0 , \quad G_{ab} \sum_{\qbf}  \frac{v(q)}{N_{\rm BZ}} q_a q_b = G_{ab}  \frac{\delta_{ab}}{2} \sum_{\qbf} \frac{v(q)}{N_{\rm BZ}} q^2 = \frac{{\rm Tr} G}{2} \sum_{\qbf} \frac{v(q)}{N_{\rm BZ}} q^2 , 
\end{equation}
for any vector $F$ and matrix $G$, which gives
\begin{equation}
\Delta_2(\psi) = \sum_{\kbf, \qbf} \frac{v(q)}{N_{\rm BZ}} \left[ |\psi_{\kbf}|^2 + \frac{q_a q_b}{8} \partial_a \partial_b |\psi_{\kbf}|^2 - |\psi_{\kbf}|^2 q_a q_b \frac{g_{\kbf_-, ab}^\uparrow + g_{\kbf_+, ab}^\downarrow}{2} - \frac{q^2}{4} |\psi_{\kbf}|^2  || S_{\kbf}^{\rm tot} ||^2 \right] . 
\end{equation}
Factorizing the sum over $\kbf$ gives an additional simplification. Indeed, the $\psi_{\kbf}$ is periodic, such that the average of any of its derivatives vanishes, and in particular 
\begin{equation}
\sum_{\kbf, \qbf} \frac{v(q)}{N_{\rm BZ}} \frac{q_a q_b}{8} \partial_a \partial_b |\psi_{\kbf}|^2 = \sum_{\qbf} \frac{q_a q_b v(q)}{8 N_{\rm BZ}} \sum_{\kbf} \partial_a \partial_b |\psi_{\kbf}|^2 = 0 .
\end{equation}
This leaves us with only three terms 
\begin{equation}
\Delta_2(\psi) = \sum_{\kbf, \qbf} \frac{v(q)}{N_{\rm BZ}} |\psi_{\kbf}|^2 \left[ 1 -  q_a q_b \frac{g_{\kbf_-, ab}^\uparrow + g_{\kbf_+, ab}^\downarrow}{2} - \frac{q^2}{4}  || S_{\kbf}^{\rm tot} ||^2 \right] , 
\end{equation}
where we have kept the quadratic term involving the quantum metric in a form similar to that of Eq.~\ref{appeq_delta1} to obtain a similar exponentiated form. 
This exponentiation is the \underline{third and last step of the calculation}, and yields
\begin{equation} \label{appeq_delta2}
\Delta_2(\psi) \approx \sum_{\kbf, \qbf} \frac{v(q)}{N_{\rm BZ}} |\psi_{\kbf}|^2  \left[ 1 - \frac{q^2}{4} |\psi_{\kbf}|^2  || S_{\kbf}^{\rm tot} ||^2 \right] e^{- \frac{1}{2} q_a (g_{\kbf_-, ab}^\uparrow + g_{\kbf_+, ab}^\downarrow) q_b} . 
\end{equation}

The main purpose of the exponential decay obtained in Eqs.~\ref{appeq_delta1} and~\ref{appeq_delta2} is to provide a characteristic cutoff and avoid divergences from large-$\qbf$. These cutoffs are approximate, and, in the limit where the spin-$\uparrow$ and spin-$\downarrow$ quantum metrics do not fluctuate too much around their mean, they are almost identical and uniform across the BZ. 
To avoid unnecessary complexity, we assume that this is the case from now on and identify 
\begin{equation}
g_{\kbf_-}^{\uparrow} \approx \frac{g_{\kbf_-, ab}^\uparrow + g_{\kbf_+, ab}^\downarrow}{2} \approx \frac{g_{\kbf, ab}^\uparrow + g_{\kbf, ab}^\downarrow}{2} = g_{\kbf} ,
\end{equation}
with the spin-averaged quantum metric introduced in the main text. 
Combining the different pieces, we finally obtain Eq.~\ref{eq_magnoninteractiongap}
\begin{equation}
\Delta(\psi) = \frac{1}{N} \sum_{\kbf} |\psi_{\kbf}|^2 \cdot || S_{\kbf}^{\rm tot} ||^2 \cdot U_{\kbf}, \quad U_{\kbf} = \left[\sum_{\qbf} \frac{v(\qbf)}{4 N_{\rm BZ}} e^{-  q_a g_{\kbf, ab} q_b} \right] . 
\end{equation}

\section{Magnon spin stiffness} \label{app_magnonspinstiffness}

In this appendix, we assume that the system has an extended spin-SU(2) symmetry and expand the topological dipole around $\Qbf=0$ to obtain the spin stiffness of the gapless magnons. 
We work in the gauge where $s_{\Qbf=0,\kbf} =1$ and $\mathcal{A}^\uparrow 
 = \mathcal{A}^\downarrow \equiv \mathcal{A}$, and find that
\begin{align}
\left( \mathcal{S}_{\Qbf,\kbf}^{\rm geom} \right)_b & = \mathcal{A}_{\kbf-\Qbf/2} - \mathcal{A}_{\kbf+\Qbf/2} + i \nabla_{\kbf} \log s_{\Qbf,\kbf} \simeq - Q_a \partial_{k_a} \mathcal{A}_{\kbf;b} + i \partial_{k_b} \log \left[1 + \frac{1}{2} Q_a (\braket{\partial_{k_a} u_{\kbf} }{u_{\kbf}} - \braket{u_{\kbf} }{\partial_{k_a} u_{\kbf}}  ) \right] \\
& = - Q_a \partial_{k_a} \mathcal{A}_{\kbf;b} + i \partial_{k_b} \log \left[1 -  Q_a \braket{u_{\kbf} }{\partial_{k_a} u_{\kbf}} \right] \simeq  Q_a \left[- \partial_{\kbf_a} \mathcal{A}_{\kbf;b} + \partial_{k_b} \mathcal{A}_{\kbf;a} \right] = Q_a \epsilon_{ba} \Omega_{\kbf} , 
\end{align}
with $\Omega_{\kbf}$ the spin-independent Berry curvature of the model. 
As a result, we get
\begin{equation}
||\mathcal{S}_{\Qbf, \kbf}^{\rm geom}||^2 \simeq |\Qbf|^2 \Omega_{\kbf}^2 , 
\end{equation} 
providing the spin stiffness quoted in the main text (Eq.~\ref{eq_minimumstiffness}). A similar expansion can be used to derive the magnons' effective mass when their spectrum is gapped.

\section{Microscopic models} \label{app_models}

 %%%%%%%%%%%%%
\begin{figure}
\centering
\includegraphics[width=0.4\columnwidth]{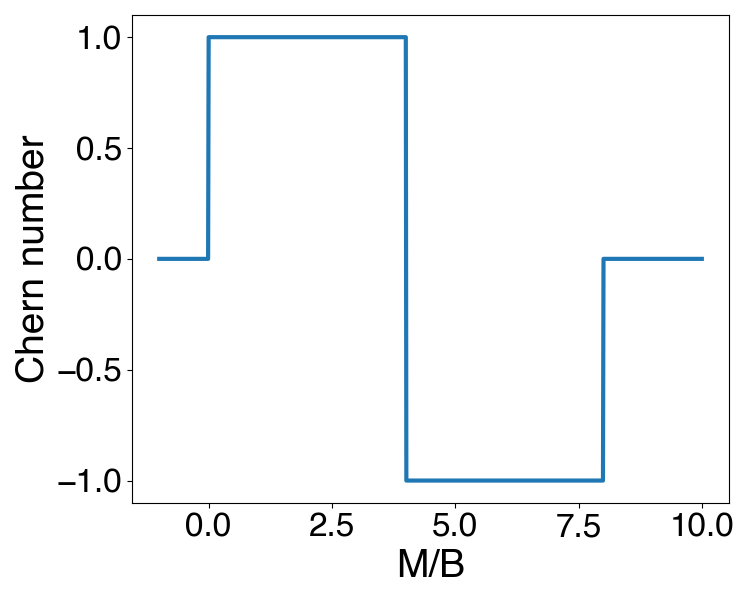}
\caption{Chern number of wavefunction $| u^{\up}_{\kbf} \rangle$ as function of $M/B$
}
\label{fig:bhz_chern}
\end{figure}
%%%%%%%%%%%%%%%%%%

\subsection{Interacting BHZ model}
\label{sm:BHZ}

We now describe the lattice Hamiltonian presented in the main text. The noninteracting Hamiltonian follows the 2d BHZ model~\cite{bernevig2006quantum}, $H_0 = {\rm diag} [H^{\up}, H^{\dn}]$, where
\begin{equation}
    H^{\sigma}(\kbf) = \begin{bmatrix}
    M - 2B(2-\cos k_x -\cos k_y) & A(\sigma \sin k_x - i\sin k_y) \\
    A(\sigma \sin k_x + i\sin k_y) & -\left[M - 2B(2-\cos k_x -\cos k_y)\right]
    \end{bmatrix}
\end{equation}
The wavefunction for $\up$ and $\dn$ spins of the lower band are
\begin{align}
    | u^{\up}_{\kbf} \rangle &= \frac{1}{\sqrt{2d_{\kbf}(d_{\kbf} - m_{\kbf} ) }} \begin{bmatrix}
    m_{\kbf} - d_{\kbf} \\
    F_{\kbf}^+
    \end{bmatrix} \\
    | u^{\dn}_{\kbf} \rangle &= \frac{1}{\sqrt{2d_{\kbf}(d_{\kbf} - m_{\kbf}) }} \begin{bmatrix} 
    m_{\kbf} - d_{\kbf} \\
    -F_{\kbf}^-
    \end{bmatrix} 
    \label{eq:BHZwavefunctions}
\end{align}
where $m_{\kbf} = M - 2B(2-\cos k_x -\cos k_y)$, $F_{\kbf}^{\pm} = A(\sin k_{x} \pm i \sin k_y$), $d_{\kbf} = \sqrt{F_{\kbf}^{+}F_{\kbf}^{-}+m_{\kbf}^2}$.  Setting $A = 1$ and $B = 1$, the Chern number as a function of $M/B$ is shown in Fig.~\ref{fig:bhz_chern}. The system exhibits two types of topological phase transitions: at $M=0$ and $M=8$, the Chern number changes from $C=0$ to $C= \pm 1$, while at $M = 4$, it switches from $C=1$ to $C=-1$.

\subsection{tMoTe$_2$ and ferromagnetism}

An extended fully polarized phase has been observed in twisted MoTe$_2$ near a twist angle of $3.7^{\circ}$~\cite{cai2023signatures}. 
In this appendix, we present the continuum model used to describe the tMoTe$_2$ system. 
In tMoTe$_2$, the low-energy valence bands predominantly originate from the $\pm K$ valleys of the monolayer. Due to strong spin–orbit coupling, the spin and valley degrees of freedom are locked to each other. 
In addition to the translational symmetry of the moir\'e lattice, the AA-stacked tMoTe$_2$ also respects $C_3$, $C_{2y}$, and time-reversal ($\mathcal{T}$) symmetries. By assuming that the top and bottom layers are rotated by $\theta/2$ and $-\theta/2$, respectively, the moir\'e lattice constant is
\begin{equation}
    a_{M} = \frac{a_{0}}{2\sin \left(\frac{\theta}{2}\right)},
\end{equation}
where $a_0 = 3.52$~\r{A} is the monolayer lattice constant. The continuous model Hamiltonian for spin-$\up$ (equivalently, at $K$ valley) takes the form as~\cite{wu2019topological,devakul2021magic,Wang2024,Jia2024}
\begin{equation}
    \mathcal{H}_{\up} = \begin{bmatrix}
        H_{t}(\rbf) & \Delta_{T}(\rbf) \\
        \Delta_{T}^{*}(\rbf) & H_{b}(\rbf)
    \end{bmatrix},
\end{equation}
where the layer Hamiltonians are given by
\begin{equation}
    H_{t/b}(\rbf) = -\frac{\hbar^2}{2m^*}(-i\nabla-\kappabf_{\pm})^2 + \Delta_{\pm}(\rbf) \pm \frac{\Delta \varepsilon}{2}.
\end{equation}
Here, $m^*=0.6 m_e$ denotes the effective electron mass, and $\Delta\varepsilon = \frac{D}{\epsilon \epsilon_0} \xi_0$ represents the potential difference induced by an applied vertical displacement field, with $\xi_0$ the interlayer distance of the MoTe$_2$ bilayer.
Due to the interlayer twist, the $K$ points of the two layers are displaced and folded into the corners of the moir\'e Brillouin zone, labeled as $\boldsymbol{\kappa}{\pm}$. 
We define the moir\'e reciprocal lattice vectors to be $\Gbf_j = \frac{4\pi}{\sqrt{3} a_M} \left(\cos \frac{\pi(j-1)}{3}, \sin \frac{\pi(j-1)}{3} \right)$, and choose $\kappabf_{+}=\frac{\Gbf_1 + \Gbf_2}{3}$, $\kappabf_{-}=\frac{\Gbf_1 + \Gbf_6}{3}$, respectively.
The intralayer moir\'e potential and the interlayer tunneling terms are given by 
\begin{equation}
    \Delta_{\pm}(\rbf) =2w_1\sum_{j=1,3,5} \cos (\Gbf_j \cdot \rbf \pm \varphi), \qquad \Delta_{T}=w_2(1+e^{-i\Gbf_2 \cdot \rbf} + e^{-i\Gbf_3 \cdot \rbf}).
\end{equation}
Because the $\pm K$ valleys are related by time-reversal symmetry, the Hamiltonian for spin-$\downarrow$ electrons ($\mathcal{H}_{\downarrow}$) can be obtained as the time-reversal conjugate of $\mathcal{H}_{\uparrow}$. Under time reversal symmetry, the wavefunction of valley-$\eta$ and layer-$l$ transforms as $\mathcal{T} c^{\dagger}_{\eta, l,\rbf}\mathcal{T} = c^{\dagger}_{-\eta, l,\rbf}$. As a result, the spin-$\dn$ Hamiltonian takes the form
\begin{equation}
    \mathcal{H}_{\dn} = \begin{bmatrix}
        H_{t}^{*}(\rbf) & \Delta_{T}^{*}(\rbf) \\
        \Delta_{T}(\rbf) & H_{b}^{*}(\rbf)
    \end{bmatrix}.
\end{equation}

For the interaction term, we consider a dual-gate screened Coulomb interaction $v(\qbf) = \frac{e^2}{2\epsilon\epsilon_0} \frac{\tanh(\xi|\qbf
|)}{|\qbf|}$, where $\xi$ is the distance from the middle of the bilayer to the gates. The parameters of the Hamiltonian we used are listed in Table.~\ref{tab:tmote2_para}. 
\label{sm:TMD}

\begin{table}
    \centering
    \begin{tabular}{c|c|c|c|c|c|c}
    \hline\hline
        $m^*$($m_e$) & $w_1 ({\rm meV})$ & $\varphi$(deg) & $w_2 ({\rm meV})$ & $\xi_0 (\rm{\r{A}})$ & $\xi$ (\r{A}) & $\epsilon$ \\
        \hline
         0.60 & 16.5 & -105.9 & -18.8 &  7 & 300 & 7\\
         \hline\hline
    \end{tabular}
    \caption{Values of the parameters for the continuum model of AA-stacking tMoTe$_2$. }
    \label{tab:tmote2_para}
\end{table}

 %%%%%%%%%%%%%
\begin{figure}
\centering
\includegraphics[width=0.8\columnwidth]{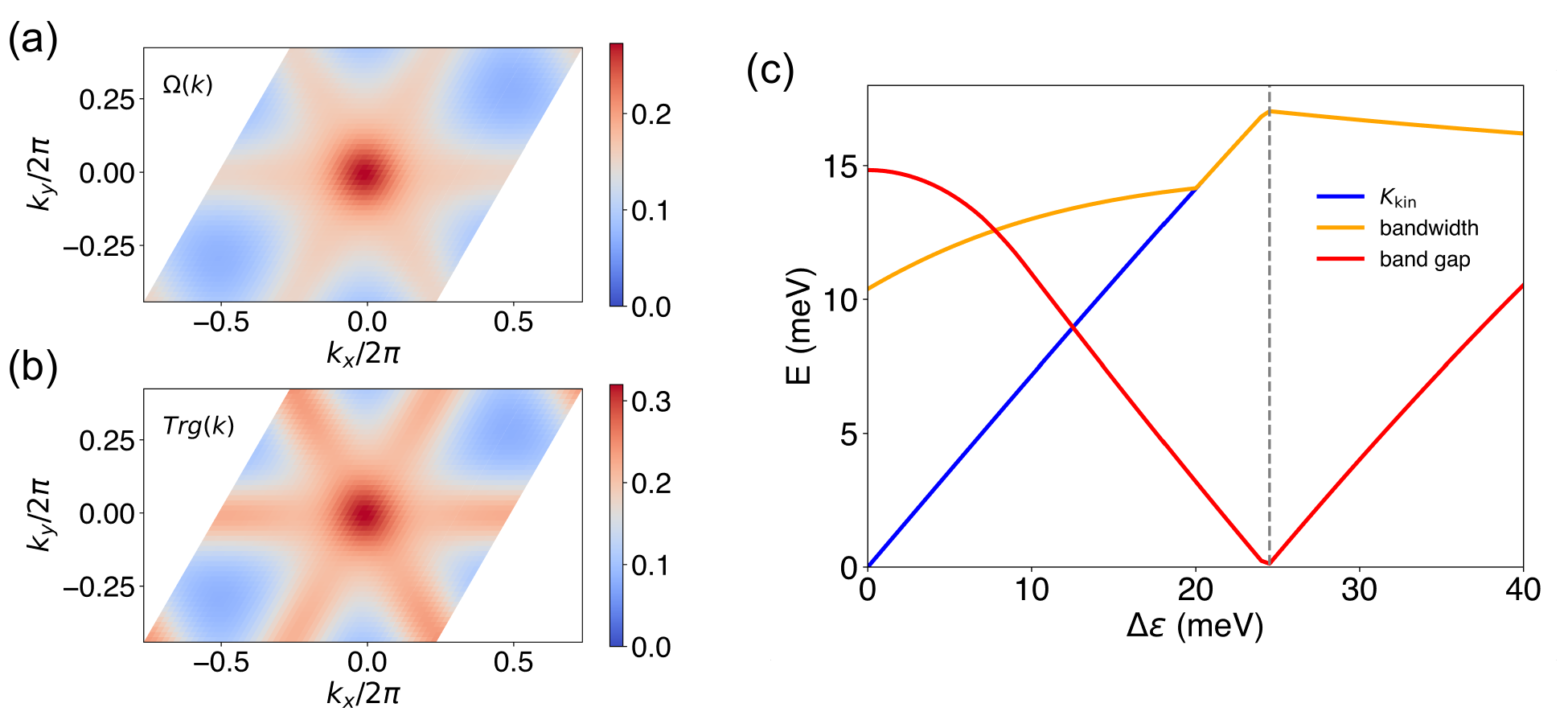}
\caption{ (a) Berry curvature $\Omega(\mathbf{k})$ and (b) trace of quantum metric $\mathrm{Tr}g(\mathbf{k})$ for spin-$\up$ with zero displacement field ($\Delta\varepsilon = 0$).
(c) Maximum energy difference $K_{\rm kin} = \max(|\varepsilon_{\kbf, \downarrow} - \varepsilon_{\kbf, \up}|)$, bandwidth of the top band, and band gap between the first and second highest bands versus the displacement potential $\Delta\varepsilon$. The gray dashed line marks the topological phase transition point.
}
\label{fig:mote2_geo}
\end{figure}
%%%%%%%%%%%%%%%%%%

Figs.~\ref{fig:mote2_geo} (a) and (b) display the Berry curvature $\Omega(\mathbf{k})$ and the trace of the quantum metric $\mathrm{Tr}g(\mathbf{k})$, respectively, for the case of zero displacement field. In Fig.~\ref{fig:mote2_geo} (c), we track how the system evolves with increasing displacement potential: the maximum energy difference between spin bands, $K_{\rm kin} = \max(|\varepsilon_{\kbf, \downarrow} - \varepsilon_{\kbf, \up}|)$, the bandwidth of the top band, and the energy gap separating the first and second highest bands are shown as a function of $\Delta\varepsilon$. The topological phase transition is marked by vertical gray dashed line with $\Delta \varepsilon_{c}=24.5~{\rm meV}$.

 %{\color{black} 
\section{Bethe–Salpeter equation in the contact-interaction limit}
\label{sm:uniform}
In this appendix we analyze the BS equation in the contact-interaction limit, where the interaction is momentum independent, $v(\qbf) = v$. In this case the interaction kernel in Eq.~\ref{eq:bs_fullform} becomes separable in the Bloch wavefunctions.

Factoring out the constant interaction kernel $v$, the diagonal interaction term in the Bethe-Salpeter Hamiltonian take the form
% \begin{equation}
% \sum_{\pbf} W_{\pbf\pbf}^{\up\up} (\kbf_{-} - \pbf) = \frac{v}{N_{\rm BZ}} \sum_{\pbf} |\langle u_{\kbf_-}^{\uparrow}|u_{\pbf}^{\uparrow} \rangle|^2 \equiv \Sigma_{\kbf}(\Qbf),
% \end{equation}
%{\color{blue}
\begin{equation}
\Sigma_{\kbf}(\Qbf) \equiv \sum_{\pbf} W_{\pbf\pbf}^{\up\up} (\kbf_{-} - \pbf) = v \bar{\rho}^\uparrow_{\kbf_-}, \quad  \bar{\rho}^\uparrow_{\kbf_-} = \braOket{u_{\kbf_-}^\uparrow}{\hat{\rho}^\uparrow}{u_{\kbf_-}^\uparrow} , \quad \hat{\rho}^\uparrow = \frac{1}{N_{\rm BZ}} \sum_{\pbf} \ket{u_{\pbf}^\uparrow} \bra{u_{\pbf}^\uparrow},
\end{equation}
%}
which acts as a momentum-dependent self-energy and scales with the local density of charge removed from the ground state (as measured by the band-projected $\hat{\rho}$). 
%{\color{blue} 
Introducing an orbital basis $\ket{a=1, \cdots, N_{\rm orb}}$ for the original model from which the topological band derives and the spin-mixing projector $P_{\kbf}(\Qbf) = \ket{u_{\kbf_-}^\uparrow} \bra{u_{\kbf_+}^\downarrow}$, we can recast the off-diagonal coefficients as  
\begin{equation}
W^{\up\dn}_{\kbf_{-},\kbf_{+}} (\kbf'-\kbf) = \frac{v}{N_{\rm BZ}} \sum_{a,b} P_{\kbf,ab}(\Qbf) P_{\kbf',ab}^*(\Qbf) , \quad P_{\kbf, ab}(\Qbf) = \braOket{a}{P_{\kbf}(\Qbf)}{b} . 
\end{equation} 
Crucially, this expression factorizes the $\kbf$ and $\kbf'$ contributions, allowing for an efficient solution of the Bethe-Salpeter equation in orbital space after summation over momenta. 
%}
%The off-diagonal term reads
% \begin{equation}
% \begin{aligned}
%     W^{\up\dn}_{\kbf_{-},\kbf_{+}} (\kbf'-\kbf) &= \frac{v}{N_{\rm BZ}} \langle u^{\up}_{\kbf'_-}| u^{\up}_{\kbf_-}\rangle \langle u^{\dn}_{\kbf_+}| u^{\dn}_{\kbf'_+}\rangle  \\
%     & = \frac{v}{N_{\rm BZ}} \sum_{ab} u^{\up*}_{\kbf'_-,a} u^{\up}_{\kbf_-,a} u^{\dn*}_{\kbf_+,b} u^{\dn}_{\kbf_+',b} \\
%     & = \frac{v}{N_{\rm BZ}} \sum_{ab}   \left(  u^{\up}_{\kbf_-,a} u^{\dn*}_{\kbf_+,b} \right) \left( u^{\up*}_{\kbf'_-,a} u^{\dn}_{\kbf_+',b} \right),
% \end{aligned}
% \end{equation}
% where $a, b$ denote orbital indices. We further define the two-particle wavefunction
% \begin{equation}
%     | \kbf, \Qbf \rangle \equiv | u_{\kbf_-}^{\uparrow}\rangle | u^{\downarrow}_{\kbf, +}\rangle^*,
% \end{equation}
% which has dimension $N^2$, if each Bloch state has dimension has dimension $N$. Its components are $P_{m,\kbf}(\Qbf) = P_{ab,\kbf}(\Qbf) = u_{\kbf_-,a}^{\uparrow} u_{\kbf_+,b}^{\downarrow*}$. The BS kernel then takes the finite-rank form
% \begin{equation}
%     \mathcal{H}_{\kbf, \kbf'}(\Qbf) =  \Sigma_{\kbf}(\Qbf)\delta_{\kbf\kbf'} - \frac{v}{N_{\rm BZ}} \sum_{m=1}^{N^2} P_{m,\kbf}(\Qbf) P^*_{m,\kbf'} (\Qbf).
% \end{equation}

Substituting into Eq.~\ref{eq:bs}, %{\color{blue} 
the Bethe-Salpeter equation becomes 
\begin{equation}
\left[ \varepsilon_{\kbf_+,\dn}-\varepsilon_{\kbf_-,\up} + v \bar{\rho}_{\kbf_-}^\uparrow -\mathcal{E} \right] \psi_{\Qbf, \kbf}  -  v \sum_{ab} P_{\kbf,ab}(\Qbf) \Psi_{ab}(\Qbf) = 0, \quad \Psi_{ab}(\Qbf) = \frac{1}{N_{\rm BZ}} \sum_{\kbf'} P_{\kbf',ab}^*(\Qbf) \psi_{\Qbf, \kbf} .
\end{equation}
Multiplying the equation by $P_{\kbf,m=cd}^*(\Qbf)$ and summing over $\kbf$ gives the following equation 
\begin{equation}
\left[ \delta_{m,n} + \Pi_{m,n} (\mathcal{E}, \Qbf) \right] \Psi_n(\Qbf) = 0, \quad \Pi_{m,n} (\mathcal{E}, \Qbf) = \frac{v}{N_{\rm BZ}} \sum_{\kbf} \frac{P_{\kbf,m}^*(\Qbf) P_{\kbf,n}(\Qbf)}{\mathcal{E} - [\varepsilon_{\kbf_+,\dn}-\varepsilon_{\kbf_-,\up} + v \bar{\rho}_{\kbf_-}^\uparrow ] }
\end{equation}
over the variable $\Psi_{n=ab}(\Qbf)$, which at fixed $\Qbf$ is a bilinear of the orbital indices and corresponds to a vector with $N_{\rm orb}^2$ independent entries. 
%}

% gives
% \begin{equation}
%     \begin{aligned}
%         \Sigma_{\kbf}(\Qbf) \psi_{\Qbf,\kbf} - \frac{v}{N_{\rm BZ}} \sum_{m} P_{m,\kbf}(\Qbf) \sum_{\kbf'}P^*_{m,\kbf'}(\Qbf) \psi_{\Qbf,\kbf'} &= \mathcal{E}(\Qbf)  \psi_{\Qbf, \kbf} \\
%         (\mathcal{E}(\Qbf) - \Sigma_{\kbf}(\Qbf)) \psi_{\Qbf,\kbf} &= -\frac{v}{N_{\rm BZ}}\sum_{m} P_{m,\kbf}(\Qbf) C_{m} (\Qbf),
%     \end{aligned}
% \end{equation}
% where
% \begin{equation}\label{eq:Cm}
%     C_{m}(\Qbf) \equiv \sum_{\kbf'} P^*_{m,\kbf'} (\Qbf) \psi_{\Qbf,\kbf'}.
% \end{equation}
% Hence the wavefunction can be expressed as \begin{equation}
%     \psi_{\Qbf, \kbf} = -\frac{v}{N_{\rm BZ}} \sum_{m=1}^{N^2} \frac{C_{m}(\Qbf)P_{m,\kbf}(\Qbf)}{\mathcal{E}_{\Qbf} - \Sigma_{\kbf}(\Qbf) }.
% \end{equation}
% Multiplying Eq.~\ref{eq:Cm} by $P_{n,\kbf}^*(\Qbf)$ and summing over $\kbf$, we have
% \begin{equation}
% \begin{aligned}
%     C_{n}(\Qbf) &= \sum_{\kbf} P^*_{n,\kbf}(\Qbf) \psi_{\Qbf,\kbf} = -\frac{v}{N_{\rm BZ}}\sum_{m} C_{m}(\Qbf) \sum_{\kbf} \frac{P^*_{n,\kbf}(\Qbf) P_{m, \kbf}(\Qbf)}{\mathcal{E}_{\Qbf} - \Sigma_{\kbf}(\Qbf)}  \\
%     & = -\frac{v}{N_{\rm BZ}} \sum_{m} C_{m} (\Qbf) \Pi_{nm}(\mathcal{E}_{\Qbf}, \Qbf)
% \end{aligned}
% \end{equation}
% where
% \begin{equation}
%     \Pi_{nm}({\mathcal{E}_{\Qbf}, \Qbf) = \sum_{\kbf} \frac{P^*_{n,\kbf}(\Qbf) P_{m, \kbf}(\Qbf)}{\mathcal{E}_{\Qbf} - \Sigma_{\kbf}(\Qbf)}}
% \end{equation}
The Bethe-Salpeter equation therefore reduces to a finite-dimensional matrix problem of size $N_{\rm orb}^2$, and its
% \begin{equation}
%     [1 + \frac{v}{N_{\rm BZ}}\Pi (\mathcal{E}_{\Qbf},\Qbf)]C(\Qbf) = 0
% \end{equation} whose 
%{\color{blue} 
eigen-energies are determined by the condition %zero mode is determined by 
% \begin{equation}
%     {\rm Det} \left[ Id_{N_{\rm orb}^2} + \frac{v}{N_{\rm BZ}}\Pi (\mathcal{E}_{\Qbf},\Qbf) \right] = 0, 
% \end{equation}
\begin{equation}
    {\rm Det} \left[ Id_{N_{\rm orb}^2} + \Pi (\mathcal{E},\Qbf) \right] = 0, 
\end{equation}
with $Id_p$ denoting the identity matrix of size $p$.
%}
Thus, the infinite momentum-space integral eigenvalue equation collapses into a finite-rank algebraic eigenvalue problem, i.e. the Sherman–Morrison reduction.
For the BHZ model, the projector $P_{\kbf}(\Qbf)$ has only 4 independent coefficients, speeding up the obtention of the spectrum (although we now need to solve a non-linear equation in the energy $\mathcal{E}$). At $\Qbf = 0$, the projector takes the explicit form 
\begin{equation}
\begin{aligned}
     P_{\kbf}(0) & = | u^{\up}_{\kbf_-}\rangle \otimes | u^{\dn}_{\kbf_+}\rangle^* \\
     & = \frac{1}{2d_{\kbf}(d_{\kbf}-m_{\kbf})} \begin{bmatrix}
         (m_{\kbf} - d_{\kbf})(m_{\kbf} - d_{\kbf}) \\
     -F_{\kbf}^{+} (m_{\kbf} -d_{\kbf}) \\
    F_{\kbf}^{+} (m_{\kbf}-d_{\kbf}) \\
    -F_{\kbf}^{+2}
     \end{bmatrix},
\end{aligned}
\end{equation}
where $d_{\kbf}, m_{\kbf}$ and $F^\pm_{\kbf}$ are defined below Eq.~\eqref{eq:BHZwavefunctions}.
For the twisted MoTe$_2$ model, in contrast, the dimension depends on the number of reciprocal lattice vectors included in the basis. In our simulations, this is typically in the few hundreds, similar to the number of discretized momenta along each direction of our lattice. In that case, the numerical efficiency should be similar for both the Bethe-Salpeter equation with discretized momenta (main text) and the Sherman-Morrison method (described in this appendix). 

% For BHZ model, at $\Qbf=0$, the component of two-particle wavefunction
% \begin{equation}
% \begin{aligned}
%      P_{\kbf}(0) & = | u^{\up}_{\kbf_-}\rangle \otimes | u^{\dn}_{\kbf_+}\rangle^* \\
%      & = \frac{1}{2d_{\kbf}(d_{\kbf}-m_{\kbf})} \begin{bmatrix}
%          (m_{\kbf} - d_{\kbf})(m_{\kbf} - d_{\kbf}) \\
%     (m_{\kbf} -d_{\kbf}) -F_{\kbf}^{\dagger} \\
%     F_{\kbf}^{+} (m_{\kbf}-d_{\kbf}) \\
%     -F_{\kbf}^{+2}
%      \end{bmatrix}
% \end{aligned}
% \end{equation}
% whose dimension is only 4. For the twisted MoTe$_2$ model, in contrast, the dimension depends on the number of reciprocal lattice vectors included in the basis.
%}

\end{document}